\documentclass[superscriptaddress,showkeys,twocolumn,nofootinbib,longbibliography]{revtex4-1}

\usepackage{amsmath}
\usepackage{amssymb}
\usepackage{graphicx}
\usepackage{epsf}
\usepackage{slashed}
\usepackage{enumitem}
\usepackage[english]{babel}
\usepackage[cp1250]{inputenc}
\usepackage{hyperref}
\usepackage{xcolor}

\usepackage[only,llbracket,rrbracket,llparenthesis,rrparenthesis]{stmaryrd} 
\usepackage{accsupp} 

\newcommand{\I}{\ensuremath{\mathrm{i}}}
\newcommand{\e}{\ensuremath{\mathrm{e}}}
\renewcommand{\d}{\ensuremath{\mathrm{d}}}
\newcommand{\cc}{\ensuremath{\mathrm{c.c.}}}

\newcommand{\abs}[1]{\left|#1\right|}

\newcommand{\refer}[1]{(\ref{#1})}

\usepackage{bbm} 

\usepackage[nodayofweek]{datetime}
\newdateformat{mydate}{\twodigit{\THEDAY}{ }\shortmonthname[\THEMONTH], \THEYEAR}
\usepackage[normalem]{ulem}

\usepackage{color}
\usepackage{ulem} 

\definecolor{myred}{rgb}{1,0,0}
\definecolor{mygreen}{rgb}{0,0.8,0.2}
\definecolor{myblue}{rgb}{0,0,1}

\definecolor{Ared}{rgb}{1,0.7,0}
\definecolor{Agreen}{rgb}{0.7,0.8,0.2}
\definecolor{Ablue}{rgb}{0,0.7,1}

\renewcommand{\emph}[1]{\textit{#1}}

\begin{document}

\definecolor{mygreen}{HTML}{006E28}
\newcommand{\kasia}[1]{{\color{mygreen}\textbf{?KS:}  #1}}
\newcommand{\trom}[1]{{\color{red}\textbf{?TR:} \color{red} #1}}
\newcommand{\filip}[1]{\textbf{?FB:} {\color{blue} #1}}
\newcommand{\aw}[1]{{\color{magenta}\textbf{?AW:}  #1}}

\title{Oscillons from $Q$-balls}

\author{F. Blaschke}
\affiliation{Research Center for Theoretical Physics and Astrophysics, Institute of Physics, Silesian University in Opava, Bezru\v{c}ovo n\'{a}m\v{e}st\'{\i}~1150/13, 746~01 Opava, Czech Republic
}
\affiliation{Institute of Experimental and Applied Physics, Czech Technical University in Prague, Husova~240/5, 110~00 Prague~1, Czech Republic}

\author{T. Roma\'{n}czukiewicz}
\affiliation{
 Institute of Theoretical Physics,  Jagiellonian University, Lojasiewicza 11, 30-348 Krak\'{o}w, Poland
}

\author{K. S\l{}awi\'{n}ska}
\affiliation{
 Institute of Theoretical Physics,  Jagiellonian University, Lojasiewicza 11, 30-348 Krak\'{o}w, Poland
}

\author{A. Wereszczy\'{n}ski}

\affiliation{
 Institute of Theoretical Physics,  Jagiellonian University, Lojasiewicza 11, 30-348 Krak\'{o}w, Poland
}
 \affiliation{Department of Applied Mathematics, University of Salamanca, Casas del Parque 2, 37008 - Salamanca, Spain
}

\affiliation{International Institute for Sustainability with Knotted Chiral Meta Matter (WPI-SKCM2), Hiroshima University, Higashi-Hiroshima, 1-3-1 Kagamiyama, Hiroshima 739-8526, Japan}

\begin{abstract}
Using Renormalization Group Theory we show that oscillons in (1+1)-dimensions can be obtained, at the leading nonlinear order, from $Q$-balls of universal complex field theories. For potentials with a nonzero cubic or quartic term the universal $Q$-ball theory is well approximated by the integrable complex sine-Gordon model. This allows us to generalize the usual perturbative expansion by Fodor et. al. beyond the simplest unmodulated oscillon case. Concretely, we explain the characteristic amplitude modulations of excited oscillons as an effect of formation of a two-$Q$-ball (two-oscillon) bound state. 
\end{abstract}

\maketitle

\section{Introduction}

Oscillons \cite{BM, G, CGM}, are spatially localized, long-living, and quasi-periodic solutions which are generic excitations in various nonlinear field theories \cite{Gleiser:2004an, Gleiser:2008ty, HS, GS-1, Am-1, Am-2, Lev, DS, Dissel, SYZ}. They found applications in cosmology, e.g., in the context of inflation \cite{G-cosm, A, FMPW, LT, Aurrekoetxea:2023jwd} or dark matter \cite{Pujo, Dark-1}, as well as  in condensed matter \cite{Charukhchyan}, where they may appear during phase transitions. Oscillons exist in fundamental theories, such as Electroweak theory \cite{Gr,Gr-2}, although their physical importance there is still unclear. 

Despite their commonness and striking similarity with (topological) solitons, they are still mysterious objects. Indeed, even their most characteristic properties, such as the unexpectedly long lifetime and modulations of the amplitude have not been explained to satisfaction. 

While topological solitons and $Q$-balls \cite{QB-1,QB-2,QB-3,QB-4} are stabilized by the corresponding topological or non-topological charges, the existence of oscillons and their unexpectedly slow decay rate into radiation has no clear reason. Although it is believed that it should be related to some hidden properties of nonlinear field equations, no single, widely accepted explanation has been given. Here $Q$-balls (via the so-called oscillon/I-ball relation \cite{K, KT, I, MT}) are considered as one possibility, where an approximately conserved $U(1)$ charge may temporarily stabilize oscillons. Another option is an approximated integrability.

Another basic and unexplained feature of oscillons is the existence of double frequency which manifests via the modulation of the amplitude. The simplest oscillons have only one, so-called fundamental, frequency with the field periodically oscillating between two maxima. However, generically, these maxima also oscillate with their own frequency. Therefore, such a modulated oscillon contains two independent degrees of freedom (DoF), which origin is also a long-standing, open problem. One reason for that is that the perturbative approach to oscillons developed by Fodor, Forgacs, Horvath, and Lukacs (FFHL) \cite{Fodor:2008es} provides an approximated treatment of unmodulated oscillons. By construction, it is based on one parameter (the fundamental frequency) and therefore it sheds no light on the problem of amplitude modulation.

Recently, it has been proposed that a modulated oscillon (excited oscillon) may be viewed as a bound state of two unmodulated, that is fundamental, oscillons \cite{Blaschke:2024uec}. In this approach, each fundamental oscillon gives one DoF and the modulation arises as an effect of mutual motion (nonlinear superposition) of the oscillons. This novel and attractive approach was, however, analyzed only for a specific field theory. Furthermore, the presented arguments used a similarity of the potential with the sine-Gordon theory.  

The present work aims to shed a new light on these crucial problems and to solve some of them.

First of all, we will show that (1+1) dimensional oscillons are indeed {\it closely related with $Q$-balls}. To obtain the connection between oscillons and $Q$-balls we apply Renormalization Group Perturbation Expansion (RGPE) that was pioneered by works of Chen, Goldenfeld, and Oono \cite{Chen:1994zza, Chen:1995ena}. Concretely, we demonstrate that oscillons can be generated from $Q$-ball solutions of some complex scalar theories. These theories enjoy a certain universality, in the sense, that at the leading nonlinearity order, many different potentials lead to the same $Q$-ball equation. In fact, we identify several {\it universality classes} which support oscillons with qualitatively distinct properties.

Importantly, the $Q$-ball theory which relates to generic oscillon models is well approximated by the {\it integrable} complex sine-Gordon theory. Exploiting this, we can consider multi-Q-ball solutions. In particular, using the two-$Q$-ball solution, we will be able to {\it analytically approximate generic modulated oscillons}. This is a step beyond the usual FFHL approach that allows us to accomplish our second goal, which is to show that a modulated oscillon is indeed a bound state of two fundamental oscillons. 

\section{Renormalization Group Perturbation Expansion}

The Renormalization Group Perturbation Expansion (RGPE) introduced in  \cite{Chen:1994zza, Chen:1995ena} is in general executed as follows:
\begin{itemize}
\item Insert a naive perturbation series into the equations of motion that correspond to small amplitude expansions in fields. The leading order equations should be, therefore, linear wave equations corresponding to field fluctuations around a chosen vacuum.
\item Insert a monochromatic wave solution as a starting point of the series with some (typically complex) \emph{bare} amplitude $A_0$.
\item Solve the naive perturbation series order-by-order until a resonant (secular) term is encountered, indicating a breakdown of the perturbation expansion at a particular \emph{cutoff} scale.
\item Redefine the bare amplitude $A_0$ in terms of a \emph{dressed} amplitude $A$ in such a way that an artificial \emph{renormalization} scale is introduced, while the cutoff scale is absorbed into the definition of $A$.
\item Derive the renormalization group equations (RGEs) as consistency conditions that the solution is independent on the renormalization scale and on the particular form of the secular terms to a given order. 
\item Solve RGEs and set the renormalization scale in such a way that all secular terms in the expansion are removed, i.e., choose a subtraction scheme. This gives a renormalized solution that is a global approximation of the true solution valid in some range of amplitudes. 
\end{itemize}

Let us illustrate the workings of the RGPE method on a very simple example and consider an anharmonic oscillator governed by the equation:
\begin{equation}
\label{eq:quartosc}
\ddot y+y+y^3 = 0 \,,
\end{equation}
where $y$ is an out-of-equilibrium distance. Inserting a naive perturbation series
\begin{equation}
y = \varepsilon y_1 +\varepsilon^2 y_2 + \varepsilon^3 y_3 + \ldots\,,
\end{equation}
where $\varepsilon$ is a book-keeping perturbation parameter, produces a sequence of equations
\begin{align}
\bigl(\partial_t^2+1\bigr) y_1 & = 0\,, \\
\bigl(\partial_t^2+1\bigr) y_2 & = 0\,, \\
\bigl(\partial_t^2+1\bigr) y_3 & = - y_1^3\,, \\
\bigl(\partial_t^2+1\bigr) y_4 & = - 3y_1^2 y_2\,, \\
 & \vdots \nonumber
\end{align}
The first order is solved as
\begin{equation}
y_1 = A_0 \e^{\I t} + \cc\,,
\end{equation}
where c.c. stands for complex conjugated terms. 

Since the non-linear corrections kick in at the third order, we can set $y_2 = 0$, while for the third order we obtain the equation:
\begin{equation}
\bigl(\partial_t^2+1\bigr) y_3 = - A_0^3 \e^{3\I t} -3 A_0 |A_0|^2  \e^{\I t} +\cc
\end{equation}
The second term on the right-hand side is a resonant term as it belongs to the kernel of the operator $\bigl(\partial_t^2+1\bigr)$. Naively, it seems that we obtain an infinite contribution to the perturbation series, i.e., $\bigl(\partial_t^2+1\bigr)^{-1}\e^{\I t} =$``$\infty$''. Thus, we are in a situation somewhat analogous to what happens at a first loop correction in most QFT calculations. In order to make progress, we need to invoke regularization. 

For such a simple case, it is enough to realize that
\begin{equation}
\frac{1}{\partial_t^2+1}\e^{\I t} = -\frac{\I}{2}(t-t_0) \e^{\I t}\,,
\end{equation}
which can be checked directly. The parameter $t_0$ is an integration constant and corresponds to the ``cutoff'' scale below which the naive expansion gives a trustable approximation. The presence of a secular term\footnote{The name itself originates from astronomy, as these terms described the long-term evolution of celestial bodies across centuries, \emph{saecula} in Latin.} $(t-t_0) \e^{\I t}$ indicates that for too large $t$ the naive perturbation series breaks down. Indeed, this sub-leading term becomes comparable to the leading $\epsilon y_1$ term when $\varepsilon^2 |t-t_0| \sim 1$.
Thus, the `bare' solution 
\begin{eqnarray}
\label{eq:sol1}
 y_{\mathrm{B}} &=&  \varepsilon A_0 \e^{\I t}+ \frac{\varepsilon^3A_0^3}{8} \e^{3\I t} \nonumber \\
 & & + \frac{3\I \varepsilon^3(t-t_0)}{2}A_0 \abs{A_0}^2\e^{\I t} + \cc\,,
\end{eqnarray}
is only reliable within a limited range around the``cutoff'' scale and cannot be regarded as a global approximation of the true solution. 

The secular term is, however, only an artefact of the perturbation series. Indeed, since the energy 
\begin{equation}
E = \frac{1}{2}\dot y^2 + \frac{1}{2}y^2 + \frac{1}{4}y^4 \,,
\end{equation}
is conserved, it follows that $y^2+ y^4/2 < 2E$ and the true solution must be bounded for all $t$. It is thus believed that were we to continue the naive perturbation series to all orders, all secular terms would necessarily sum-up into a zero.

Since the presence of the secular term is only an artefact of our expansion, we must somehow eliminate it to safe the perturbation series. This is done in two steps. First, 
we introduce the \emph{renormalization scale} $\tau$ by redefining the bare amplitude $A_0$ in terms of the dressed amplitude $A\equiv A(\tau)$ in such a way that the dependence on $t_0$ disappears: 
\begin{equation}\label{eq:baretodressed}
A_0 = A\Bigl(1+\frac{3\I \varepsilon^2 (t_0-\tau)}{2} \abs{A}^2 + \mathcal{O}(\varepsilon^3)\Bigr) \,.
\end{equation}
Thus, the solution \eqref{eq:sol1} transforms into
\begin{equation}
y = \varepsilon A\e^{\I t}+ \frac{\varepsilon^3}{8}A^3 \e^{3\I t} +\frac{3\I \varepsilon^3(t-\tau)}{2}A \abs{A}^2\e^{\I t} + \cc
\end{equation}

Since $\tau$ is an artificial parameter, we can set it to whatever we want. A shrewd option is to set $\tau = t$ since it eliminates the secular term and saves the perturbation series. 
(This choice, of course, is not unique. Nothing prevents us from choosing $\tau = t+1$. This freedom is similar to the freedom of choosing a subtraction scheme in QFT.) 
\begin{figure}
\begin{center}
\includegraphics[width=1.0\columnwidth]{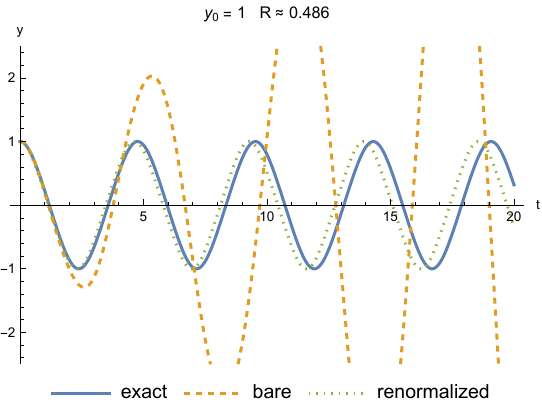}
\caption{\small Comparison among the exact (numerical) solution of Eq.~\eqref{eq:quartosc}, bare solution $y_{\mathrm{B}}$, Eq.~\eqref{eq:sol1}, and renormalized solution $y_{\mathrm{R}}$, Eq.~\eqref{eq:sol2} for $y_0 = 1$ which corresponds to $R \approx 0.486$. This choice illustrates that the match is fairly good even for a large initial amplitude. Note that for times larger than $1/R^2 \approx 4.2$  the bare solution deviates significantly from the true solution, as expected.}
\label{fig:01}
\end{center}
\end{figure}

As the second step, and 
in order to be consistent, we must demand that the solution $y$ \emph{does not depend on} $\tau$. In other words, we demand
\begin{equation}
\label{RGE}
\frac{\d y}{\d \tau} = 0
\hspace{5mm}
\mbox{for all $t$} \,.
\end{equation}
To the leading order in $\varepsilon$, the RG equation reads
\begin{equation}
\partial_\tau A = \frac{3\I \varepsilon^2}{2}A \abs{A}^2+ \mathcal{O}(\varepsilon^3) \,,
\end{equation}
with the solution
\begin{equation}
A = R\, \e^{\I \frac{3 R^2}{2}\varepsilon^2 \tau} \,,
\end{equation}
where $R$ is an arbitrary (complex) constant. 

This gives us a \emph{renormalized solution} (with $\varepsilon =1$ and $\tau = t$)
\begin{eqnarray}
y_{\mathrm{R}} &=&  2 R \cos\left[\left(\frac{3 R^2}{2} +1\right)t\right ]\nonumber \\
& & 
+ \frac{R^3}{4} \cos\left[\left(\frac{3 R^2}{2} +1\right)3t\right]\,,
\label{eq:sol2}
\end{eqnarray}
which is bounded for $t$ and which is a reliable approximation to the true solution if $|R|$ is sufficiently small.
Note, that if expanded in powers of $R$,  the renormalized solution $y_{\mathrm{R}}$ becomes equal to $y_{\rm B}$ if $A_0 = 2R$ and $\varepsilon = 1$ up to $R^3$ order. Furthermore, going from the bare to dressed amplitude \refer{eq:baretodressed} is quite analogous to wave-function renormalization in quantum field theory.

We compare the bare, renormalized, and exact (numerical) solution on Fig.~\ref{fig:01} for initial conditions $y(0) = y_0$ and $\dot y(0)=0$.

\section{Oscillons from $Q$-balls}

\subsection{A generic case}
Let us apply the RGPE algorithm to the relativistic (1+1)-dimensional scalar field theory with a generic potential 
\begin{equation}
V(\phi) = \frac{\phi^2}{2}-a_3\frac{\phi^3}{3}-a_4 \frac{\phi^4}{4}-a_5 \frac{\phi^5}{5}...,
\end{equation}
where $a_3,a_4,... \in \mathbb{R}$. Here we assumed that the oscillon oscillates around a minimum of the potential located at $\phi=0$. This can always be achieved by a target space translation. Similarly, without losing generality, the mass (curvature of the potential at the vacuum) of the field is set at $m=1$.

As before we insert into the equation of motion
\begin{equation} 
\left( \partial^2  +1 \right) \phi = a_3 \phi^2+a_4\phi^3+a_5\phi^4+... \, ,
\end{equation} 
where $\partial^2 \equiv \partial_t^2-\partial_x^2$, a naive, small-amplitude perturbation expansion, i.e.,
\begin{equation}
\phi = \varepsilon \phi_1 +\varepsilon^2 \phi_2 +\varepsilon^3 \phi_3 +\ldots  \,,
\end{equation}
where $\varepsilon$ is a book-keeping parameter.
This produces a hierarchy of equations 
\begin{subequations}
\begin{eqnarray}
\bigl(\partial^2 +1\bigr) \phi_1 & = & 0\,, \\
\bigl(\partial^2 +1\bigr) \phi_2 & = & a_3\phi_1^2\,, \\
\bigl(\partial^2 +1\bigr) \phi_3 & = & 2a_3\phi_1 \phi_2\,+a_4\phi_1^3, \\
\nonumber & \vdots &
\end{eqnarray}
\end{subequations}

Starting with a monochromatic wave at the zero order, that is
\begin{equation}
\phi_1 = A_0 \e^{\I \theta} + \cc \,,
\hspace{3mm}
\theta \equiv q x-\Omega t \,, \\
\end{equation}
where $\Omega = \sqrt{q^2+1}$, the solution up to the third order reads
{\small \begin{subequations}
\begin{eqnarray}
\phi_2 &=& -\frac{a_3}{3}A_0^2 \e^{2\I \theta}+a_3A_0 \bar A_0 + \cc\,, \\
\phi_3 &=&   \left( \frac{a_3^2}{12}-\frac{a_4}{8} \right) A_0^3 \e^{3\I \theta}+\left(\frac{10a_3^2}{3} +3a_4\right) A_0^2 \bar A_0 \mathcal{S}(\theta,\bar\theta)\e^{\I \theta} \nonumber \\ 
&\phantom{=}& +\ \cc\,,
\end{eqnarray}
\end{subequations}}
where $\bar{A}_0$ is complex conjugate of $A_0$ and we have also introduced an auxiliary variable
\begin{equation}
\bar\theta \equiv \Omega x- q t \,,
\end{equation}
and where $ \mathcal{S}(\theta,\bar\theta)$ is the secular term, i.e., a solution to the equation
\begin{equation}
\bigl(\partial^2 +1\bigr)\Bigl( \mathcal{S}(\theta,\bar\theta) \e^{\I \theta}\Bigr) =  \e^{\I \theta}\,,
\end{equation}
or, equivalently
\begin{equation}\label{eq:eqfors}
\partial_\theta^2\mathcal{S} -\partial_{\bar\theta}^2\mathcal{S} +2\I \partial_\theta \mathcal{S} = 1\,.
\end{equation}

The secular term is, of course, not unique. Given that $\mathcal{S}$ satisfies a second-order, linear PDE, we would expect it to be dependent on two arbitrary functions.
Indeed, a general solution reads
\begin{eqnarray}
\mathcal{S} &=& 
- \frac{1}{2} \bar\theta^2 + \cosh\!\Big(\bar\theta\sqrt{\partial_\theta^2+2\I \partial_\theta}\Big) g_0(\theta)
\nonumber \\ && \hspace{8.5mm} {}
+ \frac{\sinh\!\Big(\bar\theta\sqrt{\partial_\theta^2+2\I \partial_\theta}\Big)}{\sqrt{\partial_\theta^2+2\I \partial_\theta}} g_1(\theta) \,,
\end{eqnarray}
where $g_{0,1}(\theta)$ are arbitrary. Closed expressions can be obtained by choosing polynomial $g_{0,1}(\theta)$ and carrying out the derivatives.

However, physics should not depend on a particular choice of  $\mathcal{S}$, as it is just an artifact of the perturbation expansion. Indeed, we will promote this statement as a guiding principle behind how to remove the ambiguity in setting up RG equations.

Thus, let us continue to execute the RGPE algorithm with $\mathcal{S}$ being general. The bare solution up to the third order in $\varepsilon$ reads
\begin{align}
\phi_{\rm B} = &\ \varepsilon A_0 \e^{\I \theta} - \varepsilon^2\frac{a_3}{3}A_0^2 \e^{2\I \theta}+\varepsilon^2 a_3 |A_0|^2 \nonumber \\
& + \varepsilon^3   \left( \frac{a_3^2}{12}-\frac{a_4}{8} \right) A_0^3\e^{3\I\theta}   \nonumber \\
& +\varepsilon^3 \left(\frac{10a_3^2}{3} +3a_4\right) A_0 |A_0|^2\mathcal{S} \e^{\I\theta} + \cc
\end{align}
Let us now redefine the bare amplitude $A_0$ as a function of renormalized scales $\theta_0$ and $\bar \theta_0$ as
{\small \begin{equation}
A_0 = A\Big(1- \varepsilon^2 \left(\frac{10a_3^2}{3} +3a_4\right) \mathcal{S}_0(\theta_0, \bar\theta_0)|A|^2+ \mathcal{O}(\varepsilon^3) \Big)\,,
\end{equation}}
where $A \equiv A(\theta_0, \bar\theta_0)$ is the dressed amplitude.
This leads to the bare solution
\begin{align}
\phi_{\rm B} = &\ \varepsilon A \e^{\I \theta} - \varepsilon^2 \frac{a_3}{3}A^2 \e^{2\I \theta}+\varepsilon^2 a_3 |A|^2 \nonumber \\
& + \varepsilon^3   \left( \frac{a_3^2}{12}-\frac{a_4}{8} \right) A^3 \e^{3\I\theta} \nonumber \\
&  +\varepsilon^3 \left(\frac{10a_3^2}{3} +3a_4\right)  A |A|^2\bigl(\mathcal{S}-\mathcal{S}_0\bigr) \e^{\I\theta}+ \cc
\end{align}
The minimal subtraction scheme, where the entire secular term is removed once we set $\theta_0 = \theta$ and $\bar\theta_0 = \bar\theta$ is given by $\mathcal{S}_0 = \mathcal{S}$.

But first we need to ensure that the solution does not depend on the renormalization scales. 
However, demanding that
\begin{equation}
\frac{\partial \phi_{\rm B}}{\partial \theta_0} = \frac{\partial \phi_{\rm B}}{\partial \bar\theta_0} = 0 \,,
\hspace{4mm}
\forall\, x,\,t \,,
\end{equation}
leads to the RG equations in the form
\begin{subequations}
\begin{eqnarray}
\frac{\partial A}{\partial \theta_0} &=& \left(\frac{10a_3^2}{3} +3a_4\right) \varepsilon^2 A \abs{A}^2 \frac{\partial\mathcal{S}_0}{\partial\theta_0}+ \mathcal{O}(\varepsilon^3) \,,
\\
\frac{\partial A}{\partial \bar\theta_0} &=&  \left(\frac{10a_3^2}{3} +3a_4\right) \varepsilon^2 A \abs{A}^2 \frac{\partial\mathcal{S}_0}{\partial\bar\theta_0}+ \mathcal{O}(\varepsilon^3) \,,
\end{eqnarray}
\end{subequations}
that do depend on the $\mathcal{S}_0$.

Going with the minimal choice, $\mathcal{S}_0 = \mathcal{S}$, the only way how to make the RG equation independent of the form of the secular term amounts to taking a particular combination of partial derivatives, i.e.,
\begin{align}
& \ 2\I \frac{\partial A}{\partial \theta_0}+\frac{\partial^2 A}{\partial \theta_0^2}-\frac{\partial^2 A}{\partial \bar\theta_0^2} = \nonumber \\
&= \left(\frac{10a_3^2}{3} +3a_4\right)  \varepsilon^2 A |A|^2 \Bigl(2\I \frac{\partial \mathcal{S}}{\partial \theta_0} 
+\frac{\partial^2 \mathcal{S}}{\partial \theta_0^2}   -\frac{\partial^2 \mathcal{S}}{\partial \bar\theta_0^2}\Bigr)  \nonumber \\ 
& =  \left(\frac{10a_3^2}{3} +3a_4\right)  \varepsilon^2  A |A|^2\,, \hspace{4mm}
\forall\, \theta_0,\,\bar\theta_0,\, \mathcal{S} \,,
\end{align}
where in the last line we used Eq.~(\ref{eq:eqfors}). In other words, the RG equation reads
\begin{equation}\label{eq:rg}
2\I \frac{\partial A}{\partial \theta_0}+\frac{\partial^2 A}{\partial \theta_0^2}-\frac{\partial^2 A}{\partial \bar\theta_0^2} =
 \left(\frac{10a_3^2}{3} +3a_4\right) \varepsilon^2 A |A|^2\,. 
\end{equation}
This, in turn, gives us a \emph{renormalized} solution
\begin{align}\label{eq:renormsol}
\phi_{\rm R} =&\ \varepsilon A \e^{\I \theta} - \varepsilon^2 \frac{a_3}{3}A^2 \e^{2\I \theta}+\varepsilon^2 a_3 |A|^2 \nonumber \\
& + \varepsilon^3   \left( \frac{a_3^2}{12}-\frac{a_4}{8} \right) A^3 \e^{3\I\theta} + \cc
\end{align}

The RG equation \refer{eq:rg} can be further simplified by defining a new complex field
\begin{equation}
\Psi \equiv \varepsilon \sqrt{\frac{5a_3^2}{3} +\frac{3a_4}{2} } A \e^{\I \theta}\,,
\end{equation}
for which the RG equation takes the form
\begin{equation}\label{eq:RGrel}
\partial^2 \Psi +\Psi = 2 \Psi |\Psi|^2\,.
\end{equation}
This equation follows from the Lagrangian 
\begin{equation}\label{eq:lag1}
\mathcal{L} = |\partial \Psi |^2 -|\Psi |^2 +|\Psi |^4\,,
\end{equation}
that describes a relativistic field theory with an upside-down wine-bottle potential supporting $Q$-balls as spatially localized, periodic solutions. 

Indeed, taking $\Omega = 1$ and $q=0$ so that $\theta = t$ and $\bar\theta =x$, we find a stationary $Q$-ball solution:
\begin{equation}\label{eq:qball}
\Psi = \frac{\lambda }{\cosh(\lambda x)}\e^{\I \omega t}\,,
\end{equation}
where $\lambda$ is a scale parameter and $\omega = \sqrt{1-\lambda^2}$ is the $Q$-ball's frequency.

It can be easily shown by applying the RGPE method on the RG equation \refer{eq:RGrel} itself that this equation is \emph{a fixed point} of the RG flow. In other words, the condition that allows independence on the secular term would give us Eq.~\refer{eq:RGrel} itself.

In terms of $\Psi$, the renormalized solution takes a particularly simple {\it dressed} form
\begin{align}
\phi_{\rm R} = & \  \left(\frac{5a_3^2}{3} +\frac{3a_4}{2}\right)^{-1/2}  \Psi  \nonumber \\
& -\frac{a_3}{5a_3^2+\frac{9}{2}a_4 }\Psi^2 +\frac{3a_3}{5a_3^2+\frac{9}{2}a_4 }  |\Psi|^2 \nonumber \\ \label{eq:renormsol2}
&+  \left( \frac{a_3^2}{12}-\frac{a_4}{8} \right) \left(\frac{5a_3^2}{3} +\frac{3a_4}{2}\right)^{-3/2}   \Psi^3 +\cc
\end{align}

The main result is that oscillons in a generic (1+1) dimensional model emerge from a complex scalar field theory with $Q$-balls being seeds for the oscillons. Hence, there is indeed an intimate relationship between the oscillons and $Q$-balls. Importantly, the $Q$-ball theory is -- for a generic potential -- given by the same, universal Lagrangian \refer{eq:lag1}, which does not depend on the details of the original oscillon model (at least not to leading order in the perturbation expansion). On the other hand, a map, i.e., {\it dressing formula}, relating the $Q$-ball solution with the oscillon includes coefficients of the potential. 

It is important to note that the renormalized solution \refer{eq:renormsol} or \refer{eq:renormsol2} only exists if
\begin{equation}\label{eq:ourcond}
\frac{10}{3}a_3^2+3 a_4  >0,
\end{equation}
which is consistent with the condition given in \cite{Fodor:2008es} for the existence of oscillons.

\subsection{Some non-generic cases}

The renormalized solution \refer{eq:renormsol} was derived under the tacit assumption that at least one of the coefficients $a_3$ and $a_4$ is non-zero. Furthermore, if the coefficients saturate the condition \refer{eq:ourcond}, i.e., $10 a_3^2+9a_4 = 0$, the RGPE algorithm is also valid, (or, indeed, mute) as the resonant term disappears and we need to proceed to higher orders in the perturbation series.

Let us now comment on such non-generic situations.
To begin with, let us execute the RGPE algorithm for potentials that have $a_3 = a_4 =0$, i.e.,
\begin{equation}
V=\frac{\phi^2}{2}-\frac{a_5}{5}\phi^5-\frac{a_6}{6}\phi^6 - \ldots
\end{equation}
The equation of motion is
\begin{equation}
\bigl(\partial^2+1\bigr)\phi = a_5\phi^4+ a_6 \phi^5 + \ldots
\end{equation}
The naive perturbation expansion (where we skip powers of $\varepsilon$ that gives zero contribution) 
\begin{equation}
\phi = \varepsilon \phi_1 + \varepsilon^4 \phi_4 + \varepsilon^5 \phi_5 +\ldots 
\end{equation}
leads to the hierarchy of equations
\begin{align}
\bigl(\partial^2 +1 \bigr)\phi_1 & = 0\,, \\
\bigl(\partial^2 +1 \bigr)\phi_4 & = a_5\phi_1^4\,, \\
\bigl(\partial^2 +1 \bigr)\phi_5 & = a_6\phi_1^5 \,, \\
& \vdots \nonumber 
\end{align}
Plugging in  $\phi_1 = A_0 \e^{\I \theta}+ \cc$, the higher-order corrections read
{\small \begin{align}
\phi_4 & = -\frac{a_5 A_0^4}{15}\e^{4\I \theta} -\frac{4 a_5 A_0^2|A_0|^2}{3}\e^{2\I \theta} + 3a_5 |A_0|^4 + \cc\,, \\
\phi_5 & =  - \frac{a_6 A_0^5}{24}\e^{5\I \theta}-\frac{5 a_6 A_0^3|A_0|^2}{8}\e^{3\I \theta}+10a_6 A_0 |A_0|^4 \e^{\I \theta} \mathcal{S} \\
\nonumber& \phantom{=} + \cc
\end{align}}
Note that only $\phi_5$ contains the secular term.
The algorithm now calls for a redefinition of bare amplitude in terms of dressed amplitude:
\begin{equation}
A_0 = A \Bigl(1-10 \varepsilon^4 |A|^4 a_6 \mathcal{S}_0\Bigr) + \mathcal{O}\bigl(\varepsilon^5\bigr)\,.
\end{equation}
Going with the minimal subtraction scheme $\mathcal{S}_0 = \mathcal{S}$, the RG equations for the dressed amplitude reads
{\small \begin{equation}
2\I \partial_{\theta_0} A +\partial^2 A = 10\varepsilon^4 a_6 A |A|^4 + \mathcal{O}\bigl(\varepsilon^5\bigr)\,, \quad \forall \theta_0,\, \bar\theta_0,\, \mathcal{S}\,.
\end{equation}}

Defining a complex field
\begin{equation}
\Psi \equiv \varepsilon A \e^{\I t}/\gamma\,, \quad \gamma \equiv \sqrt[4]{\frac{3}{10\, a_6}}\,,
\end{equation}
the RG equations attain the following form
\begin{equation}\label{eq:rgn2}
\bigl(\partial^2+1\bigr)\Psi = 3 \Psi |\Psi|^4\,.
\end{equation}
This differs significantly from the previous RG equation which had a cubic nonlinearity. Consequently, the $Q$-ball solution is also quite different 
\begin{equation}\label{eq:Qball56}
\Psi = \frac{\sqrt{\lambda}}{\sqrt{\cosh(2\lambda x)}}\e^{\I \sqrt{1-\lambda^2}t}\,.
\end{equation}
The Lagrangian that yield this RG equation is given as
\begin{equation}
\mathcal{L} = \partial_\mu \bar\Psi \partial^\mu \Psi -|\Psi|^2 +|\Psi|^6,
\end{equation}
where, instead of a quartic self-interaction term, we have a sextic term. As we already remarked, we would have obtained the same RG equation for the models with $10 a_3^2+9a_4=0$.

The renormalized solution is now given as
\begin{align}\label{eq:renorm56}
\phi_{\rm R} = &\,  \gamma \Psi- \frac{a_5\gamma^4}{15}\Psi^4-\frac{4a_5 \gamma^4}{3}\Psi^2 |\Psi|^2+3a_5\gamma^4 |\Psi|^4 \nonumber \\
& - \frac{a_6 \gamma^5}{24}\Psi^5 -\frac{5a_6 \gamma^5}{8}\Psi^3|\Psi|^2 + \cc
\end{align}
The fact that the $Q$-ball Lagrangian has a very different form may suggest that oscillons in field theories of this type belong to a different universality class. We will analyze this problem in the next sections. 

An even more drastic departure from the complex $|\Psi|^4$ theory for the underlying $Q$-balls is found for the potential with $a_6=0$. 
For concreteness, let us consider a simple potential with quintic non-linearity:
\begin{equation}
V=\frac{\phi^2}{2}-\frac{\phi^5}{5}\,,
\end{equation}
where we set $a_5=1$ without loss of generality.
 
Performing our perturbative scheme we find the following equations
\begin{subequations}
\begin{eqnarray}
\bigl(\partial^2 +1\bigr) \phi_1 & = & 0\,, \\
\bigl(\partial^2 +1\bigr) \phi_4 & = & \phi_1^4\,, \\
\bigl(\partial^2 +1\bigr) \phi_7 & = & 4\phi_1^3 \phi_4\,, \\
\nonumber & \vdots &
\end{eqnarray}
\end{subequations}
where we omit all terms containing $\phi_2,\phi_3,\phi_5,\phi_6$ which can be consistently set to zero. The nontrivial contributions are 
\begin{align}
\phi_1 = &\  A_0 \e^{\I \theta} + \cc \\
\phi_4 = & - \frac{1}{15}A_0^4 \e^{4\I \theta}-  \frac{4}{3}A_0^2 |A_0|^2 \e^{2\I \theta}+ 3 |A_0|^4 + \cc \\
\phi_7 = &  \frac{1}{180}  A_0^7\e^{7\I\theta} +  \frac{29}{90}  A_0^5 |A_0|^2 \e^{5\I\theta}   \nonumber \\
& -  \frac{9}{10}  A_0^3 |A_0|^4 \e^{3\I\theta} +   \frac{252}{5}  A_0 |A_0|^6   \mathcal{S} \e^{\I\theta} + \cc \, ,
\end{align}
where the secular term obeys the same differential equation (\ref{eq:eqfors}). 

As before we define a dressed amplitude $A(\theta_0,\bar{\theta}_0)$ in terms of renormalized scales $\theta_0$ and $\bar \theta_0$ as
\begin{equation}
A_0 = A\Big(1- \varepsilon^6   \frac{252}{5} \mathcal{S}_0(\theta_0, \bar\theta_0)|A|^6+ \mathcal{O}(\varepsilon^7) \Big)\,.
\end{equation}
Applying the minimal subtraction scheme we arrive at the following RG equations 
\begin{equation}
 2\I \frac{\partial A}{\partial \theta_0}+\frac{\partial^2 A}{\partial \theta_0^2}-\frac{\partial^2 A}{\partial \bar\theta_0^2} =  \frac{252}{5} \varepsilon^6 A \abs{A}^6\,, \hspace{4mm}
\forall\, x,\,t,\, \mathcal{S} \,.
\end{equation}
Defining a new field
\begin{equation}
\Psi = \epsilon \left( \frac{252}{20} \right)^{1/6} A e^{\I\theta}
\end{equation}
we bring it to a  simple complex equation
\begin{equation}\label{eq:rgn3}
\partial^2 \Psi +\Psi= 4\Psi |\Psi|^6.
\end{equation}
This again differs significantly from the previous two RG equations. The $Q$-ball solution reads 
\begin{equation}
\Psi= \sqrt[3]{\frac{\lambda }{\cosh(3\lambda x)}} e^{\I \sqrt{1-\lambda^2} t}\,.
\end{equation}

In general, the potential of the form
\begin{equation}
V = \frac{1}{2}\phi^2 - \frac{1}{2(n+1)}\phi^{2(n+1)}\,,
\end{equation}
leads to a RG equation
\begin{equation}
\bigr(\partial^2+1\bigl)\Psi = (n+1)\Psi |\Psi|^{2n}\,,
\end{equation}
which has a  $Q$-ball solution
\begin{equation}
\Psi = \Bigl(\frac{\lambda}{\cosh(n \lambda x)}\Bigr)^{\tfrac{1}{n}}\e^{\I \sqrt{1-\lambda^2}t}\,.
\end{equation}

\section{Unmodulated oscillons}

\begin{figure}
\begin{center}
\includegraphics[width=1.0\columnwidth]{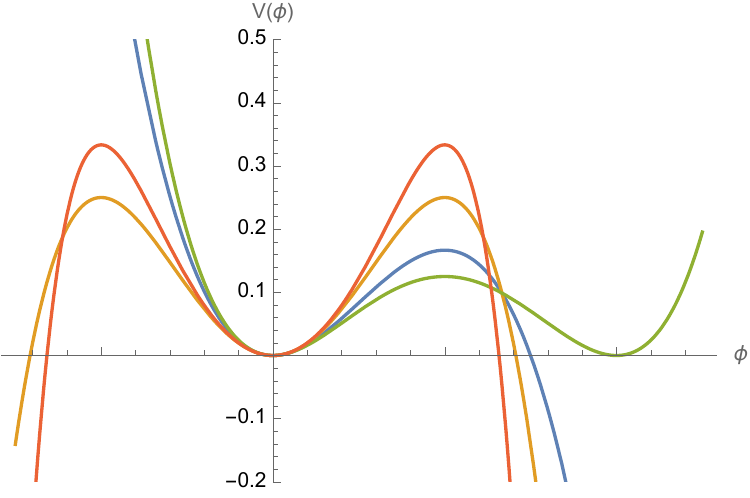}
\caption{\small Field theoretical potentials in models of oscillons that are discussed in our examples. Blue: $\phi^3$; orange: the inverse $\phi^4$; green: the double well $\phi^4$; red: the exotic $\phi^6$. }
\label{fig:pot}
\end{center}
\end{figure}

In this section, we will compare oscillons derived from our approximate scheme with the true, numerically computed solutions. We will analyze several field theories at various regimes. Specifically, we will focus on the $\phi^3$, the reverse $\phi^4$, the double well $\phi^4$ and the exotic $\phi^6$ models, see Fig. \ref{fig:pot}. The main result is that oscillons generated from a single $Q$-ball approximate unmodulated oscillons very well. These are typically small amplitude oscillons.

Furthermore, we will show that the oscillons of the first three models differ from the fourth one. They are of a different universality class as they come from the different  $Q$-ball equation. 

\subsection{$\phi^3$ theory}
\begin{figure*}
\begin{center}
\includegraphics[width=1.0\columnwidth]{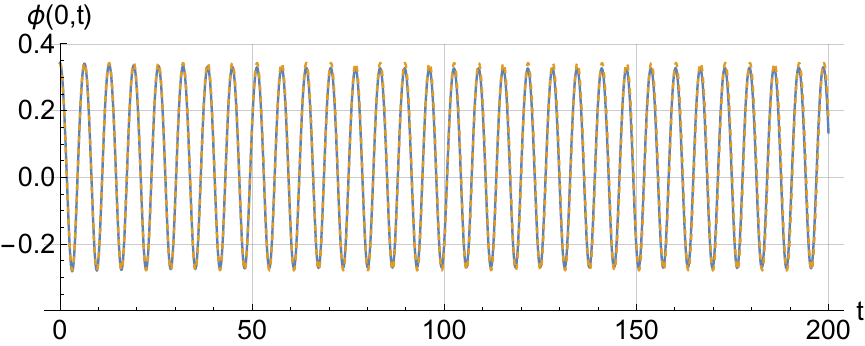}
\includegraphics[width=1.0\columnwidth]{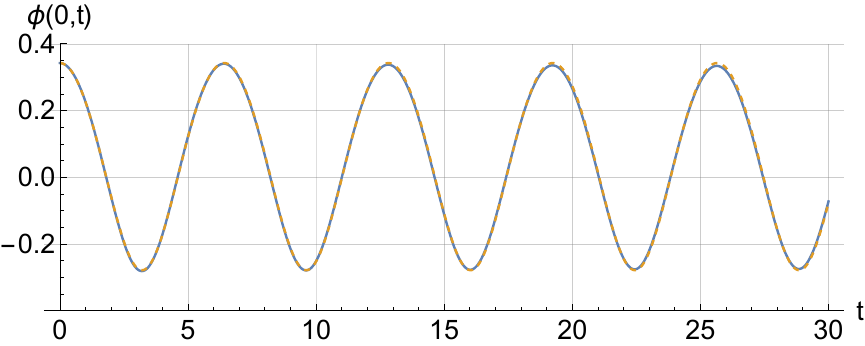}
\includegraphics[width=1.0\columnwidth]{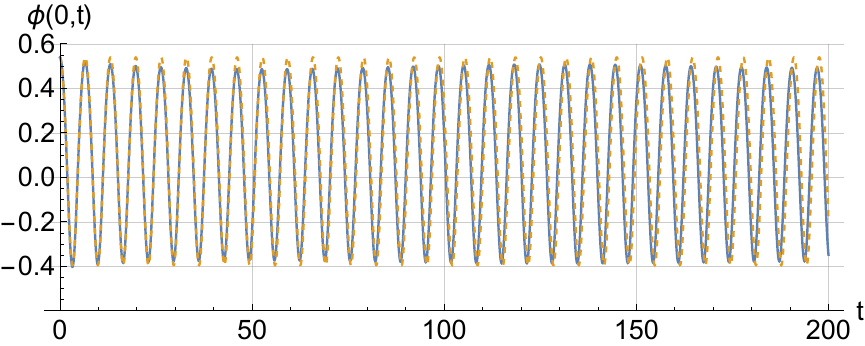}
\includegraphics[width=1.0\columnwidth]{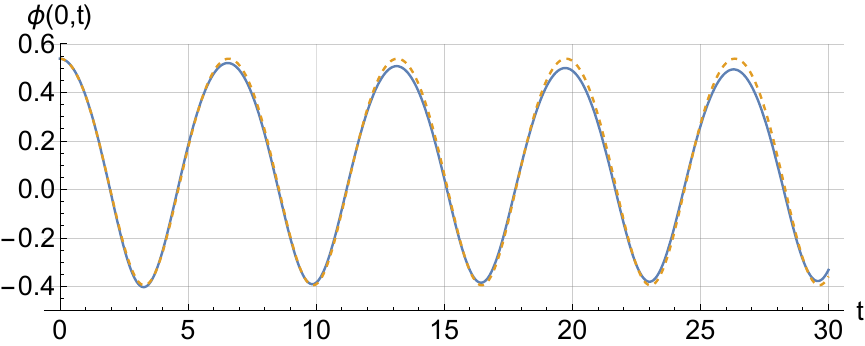}
\includegraphics[width=1.0\columnwidth]{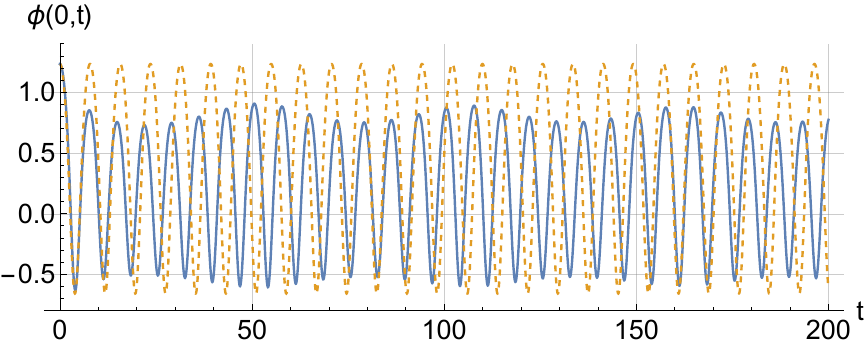}
\includegraphics[width=1.0\columnwidth]{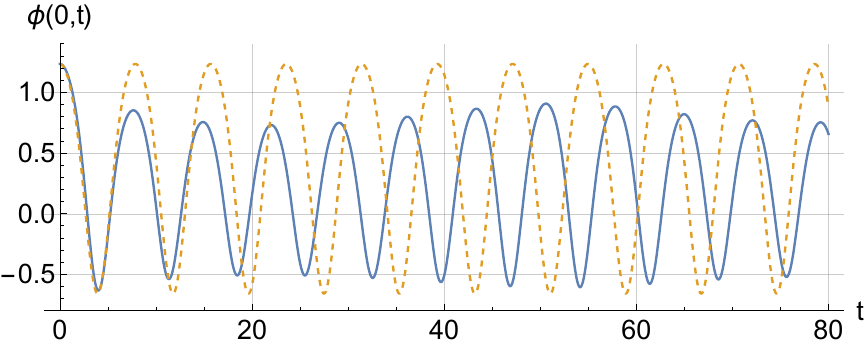}
\caption{\small Comparison between numerically found oscillon (blue) and renormalized solution (orange) for the single Q-ball solution in $\phi^3$ theory. We plot the value of the field at origin $\phi(x=0,t)$. Upper: $\lambda = 0.2$; Lower: $\lambda= 0.3$; Bottom: $\lambda= 0.6$. }
\label{fig:phi3}
\vspace*{0.5cm}
\includegraphics[width=0.5\columnwidth]{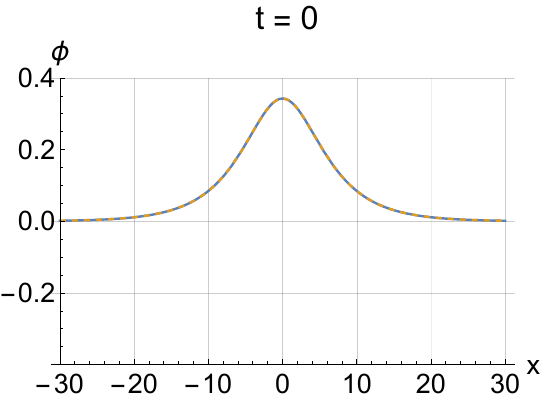}
\includegraphics[width=0.5\columnwidth]{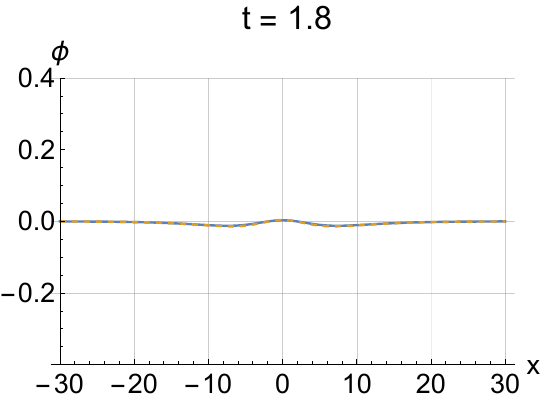}
\includegraphics[width=0.5\columnwidth]{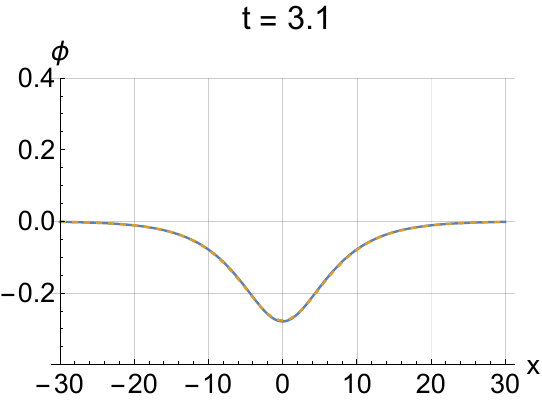}
\includegraphics[width=0.5\columnwidth]{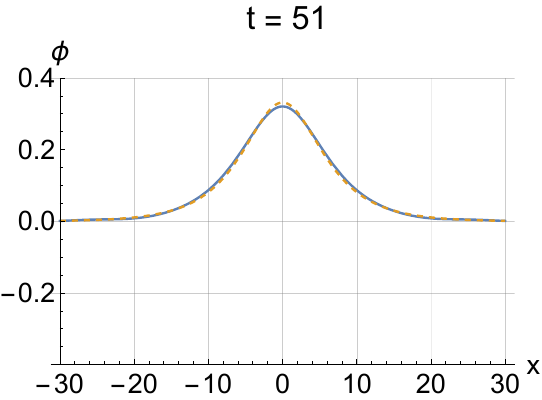}

\includegraphics[width=0.5\columnwidth]{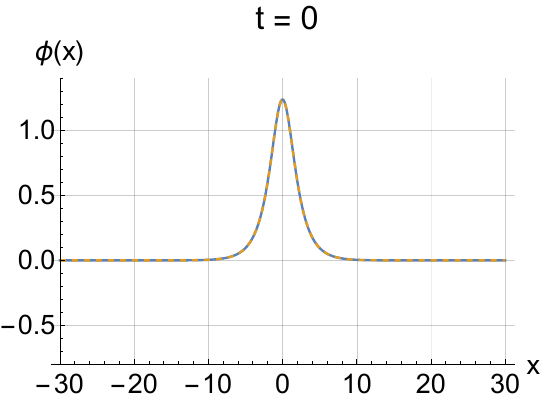}
\includegraphics[width=0.5\columnwidth]{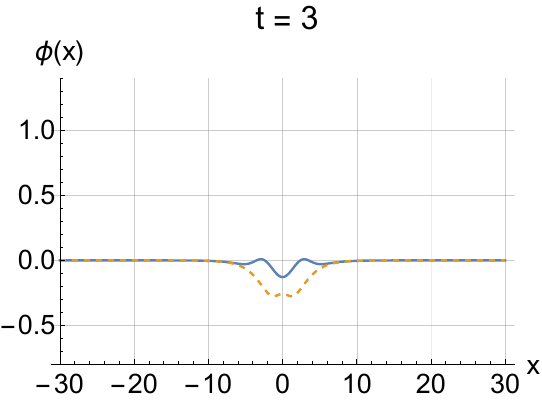}
\includegraphics[width=0.5\columnwidth]{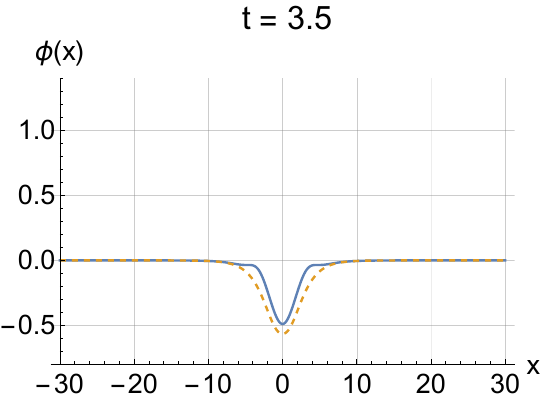}
\includegraphics[width=0.5\columnwidth]{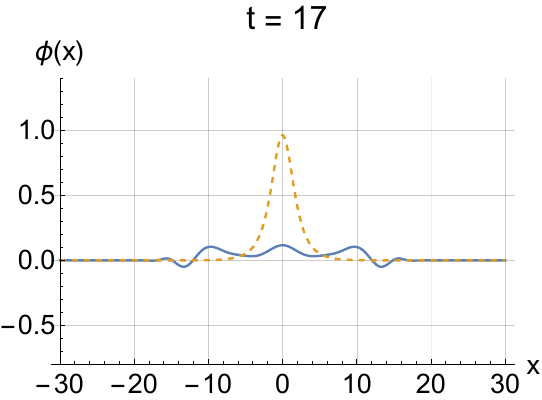}
\caption{\small Comparison between numerically found oscillon (blue) and renormalized solution (orange) for the single Q-ball solution in $\phi^3$ theory. Upper: $\lambda=0.2$ and we plot the field profiles at $t=0, 1.8, 3.1$ and $t=51$.  Lower: $\lambda=0.6$ and $t=0, 3,3.5$ and $t=17$.}
\label{fig:phi3-profile}
\end{center}
\end{figure*}
\begin{figure*}
\begin{center}
\includegraphics[width=1.0\columnwidth]{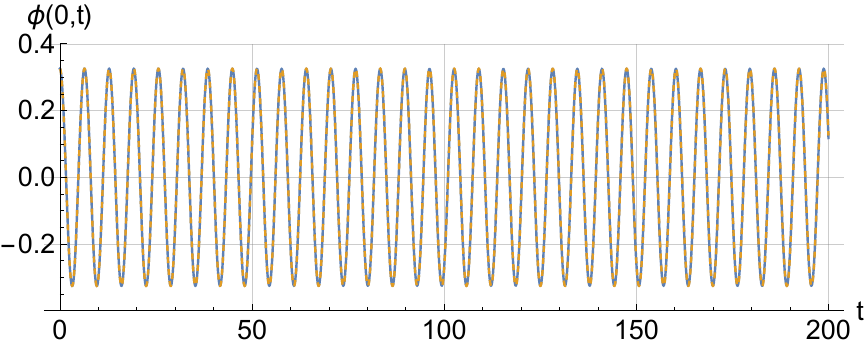}
\includegraphics[width=1.0\columnwidth]{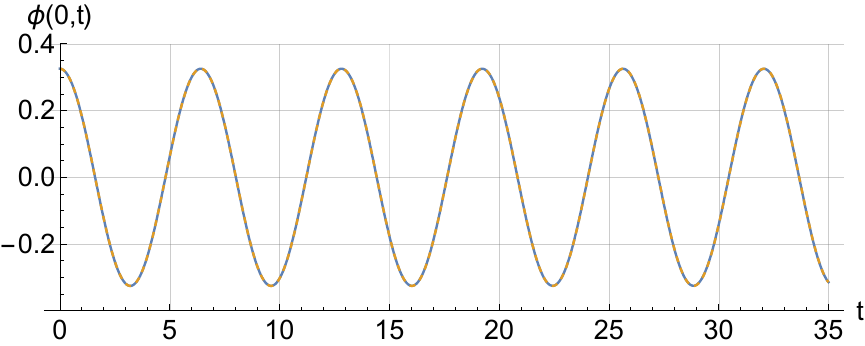}
\includegraphics[width=1.0\columnwidth]{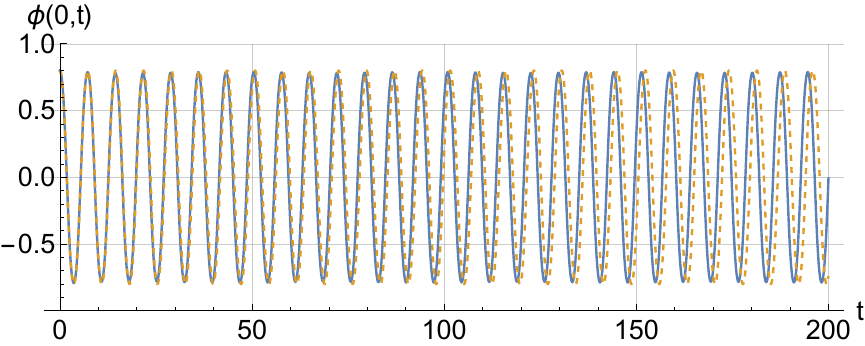}
\includegraphics[width=1.0\columnwidth]{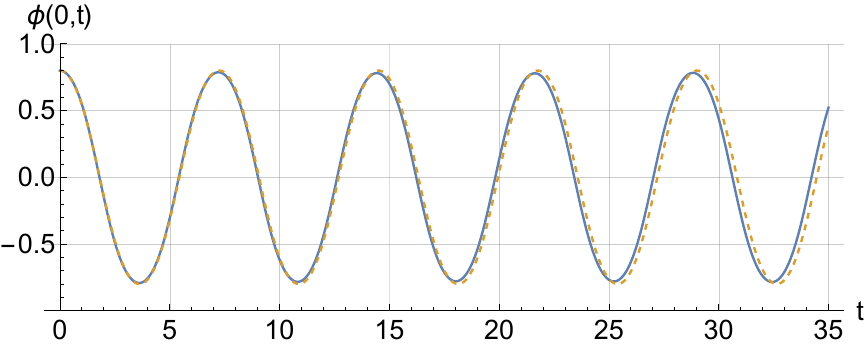}
\includegraphics[width=1.0\columnwidth]{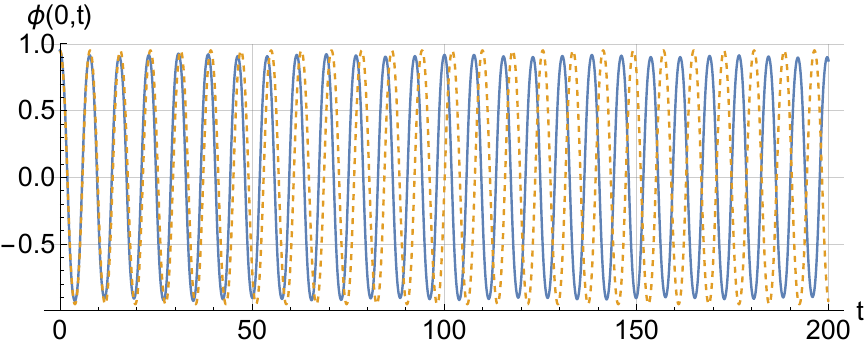}
\includegraphics[width=1.0\columnwidth]{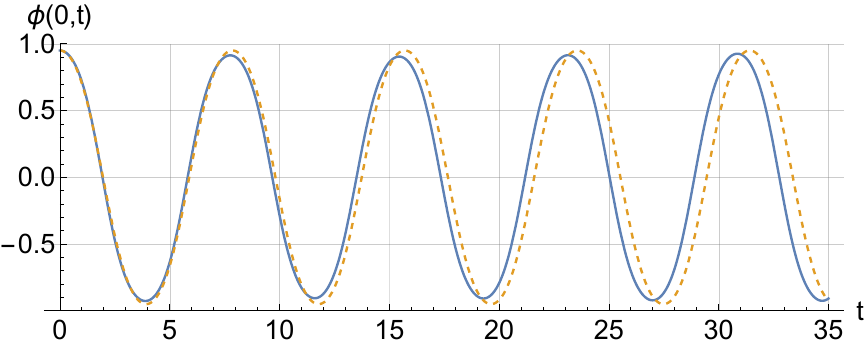}
\caption{\small Comparison between numerically found oscillon (blue) and renormalized solution (orange) for the single Q-ball solution in the inverse $\phi^4$ theory. We plot the value of the field at origin $\phi(x=0,t)$. Upper: $\lambda = 0.2$; Lower: $\lambda= 0.5$; Bottom: $\lambda= 0.6$. }
\label{fig:phi4}
\vspace*{0.5cm}
\includegraphics[width=0.5\columnwidth]{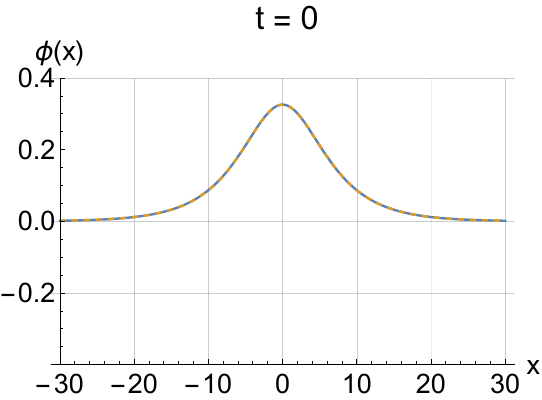}
\includegraphics[width=0.5\columnwidth]{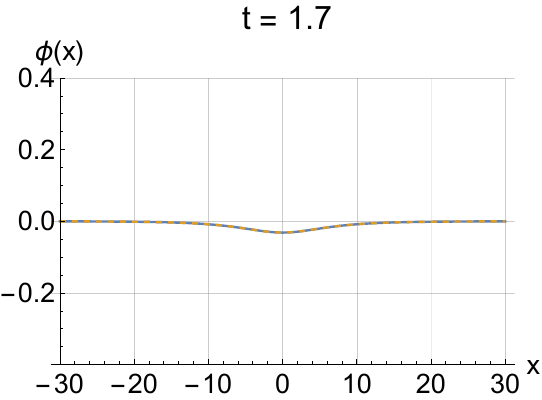}
\includegraphics[width=0.5\columnwidth]{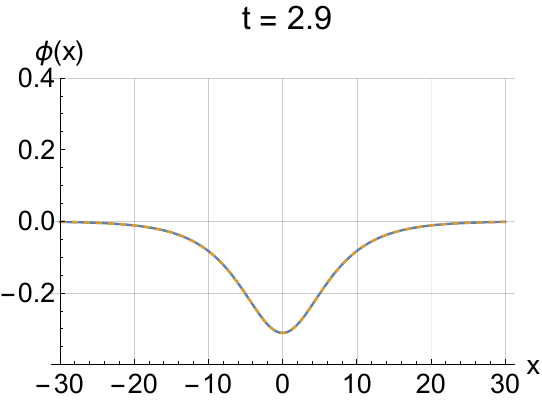}
\includegraphics[width=0.5\columnwidth]{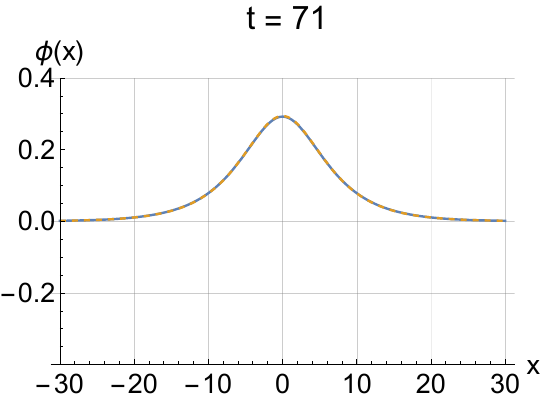}

\includegraphics[width=0.5\columnwidth]{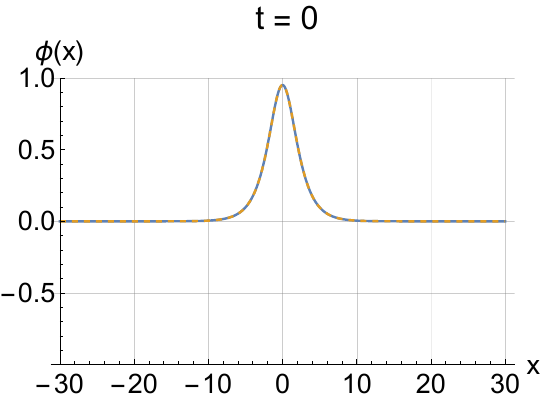}
\includegraphics[width=0.5\columnwidth]{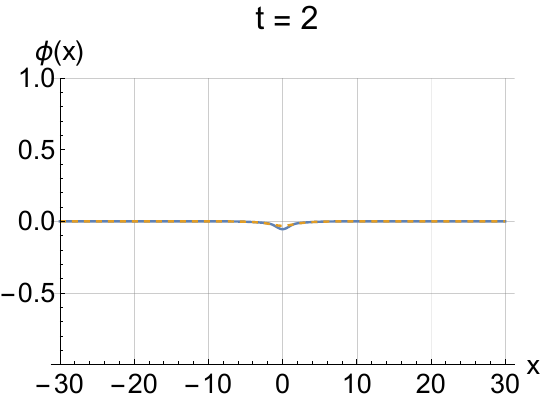}
\includegraphics[width=0.5\columnwidth]{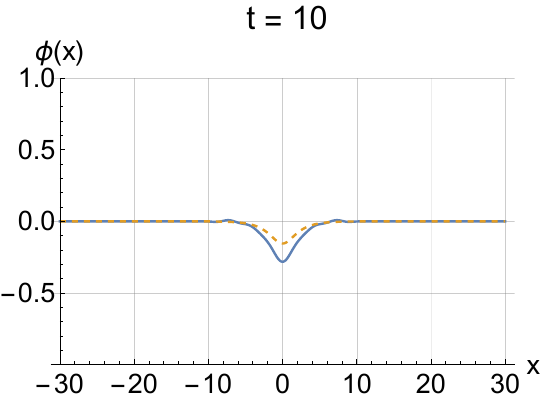}
\includegraphics[width=0.5\columnwidth]{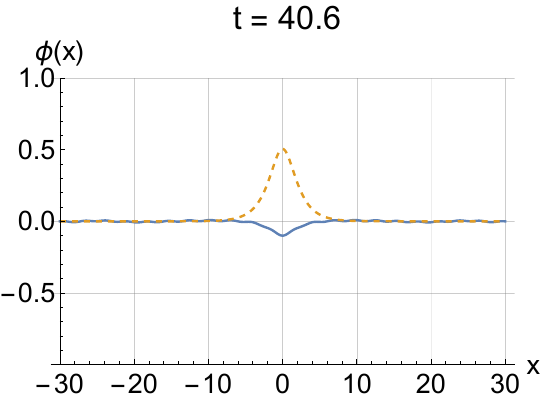}
\caption{\small Comparison between numerically found oscillon (blue) and renormalized solution (orange) for the single Q-ball solution in the inverse $\phi^4$ theory. Upper: $\lambda=0.2$ and we plot the field profiles at $t=0, 1.7, 2.9$ and $t=71$.  Lower: $\lambda=0.6$ and $t=0, 2,10$ and $t=40.6$.}
\label{fig:phi4-profile}
\end{center}
\end{figure*}

\begin{figure*}
\begin{center}
\includegraphics[width=1.0\columnwidth]{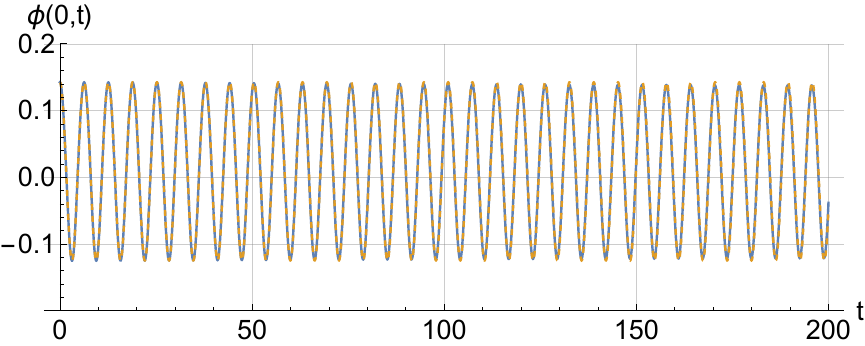}
\includegraphics[width=1.0\columnwidth]{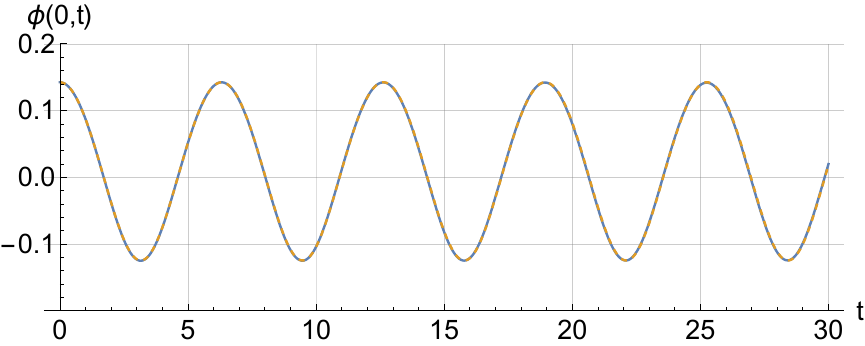}
\includegraphics[width=1.0\columnwidth]{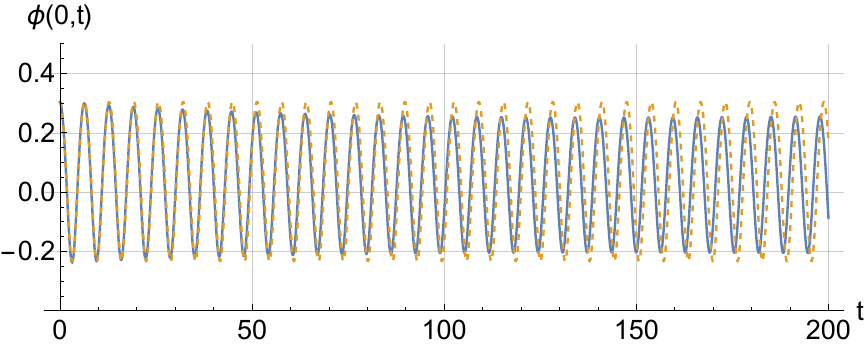}
\includegraphics[width=1.0\columnwidth]{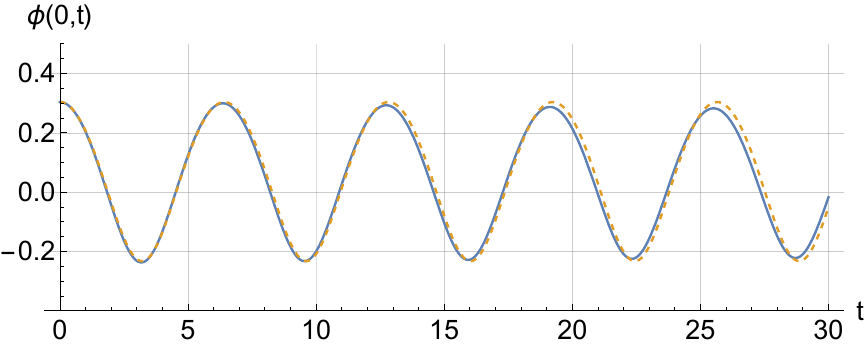}
\includegraphics[width=1.0\columnwidth]{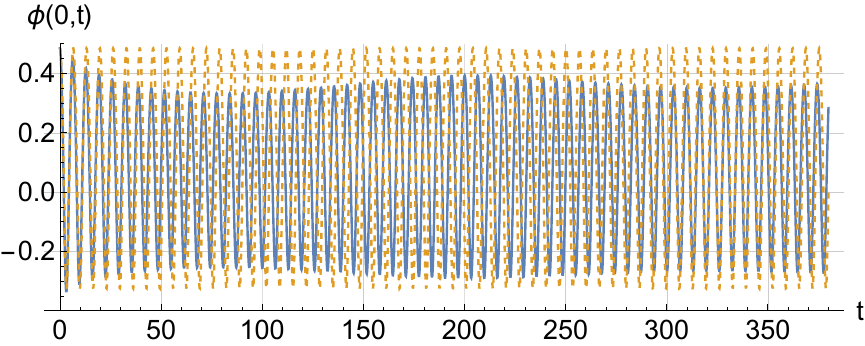}
\includegraphics[width=1.0\columnwidth]{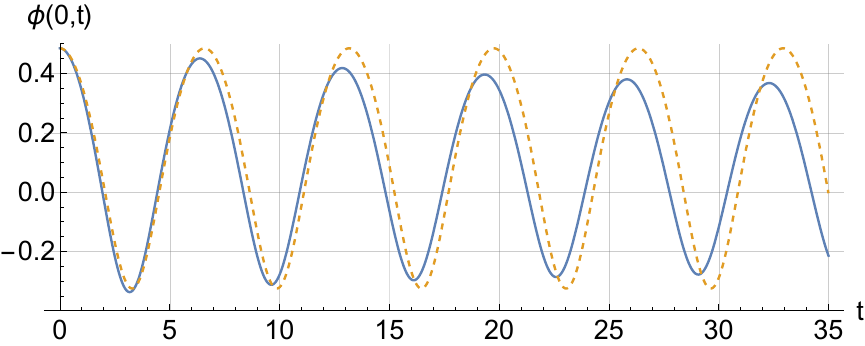}
\caption{\small Comparison between numerically found oscillon (blue) and renormalized solution (orange) for the single Q-ball solution in the double well $\phi^4$ theory. We plot the value of the field at origin $\phi(x=0,t)$. Upper: $\lambda = 0.1$; Lower: $\lambda= 0.2$; Bottom: $\lambda= 0.3$. }
\label{fig:dw}
\vspace*{0.5cm}
\includegraphics[width=0.5\columnwidth]{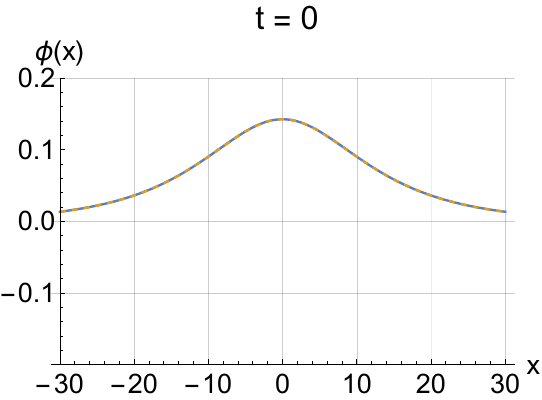}
\includegraphics[width=0.5\columnwidth]{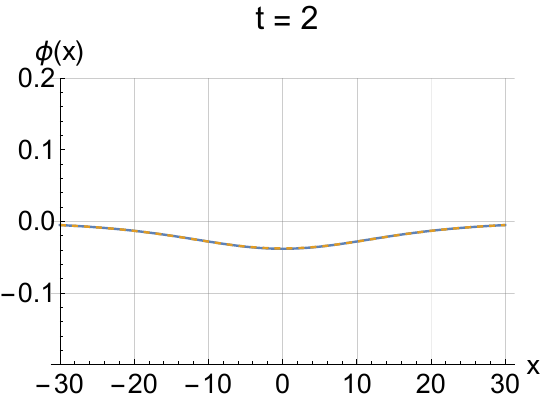}
\includegraphics[width=0.5\columnwidth]{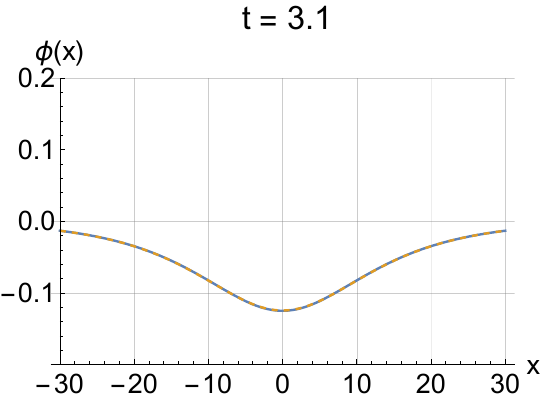}
\includegraphics[width=0.5\columnwidth]{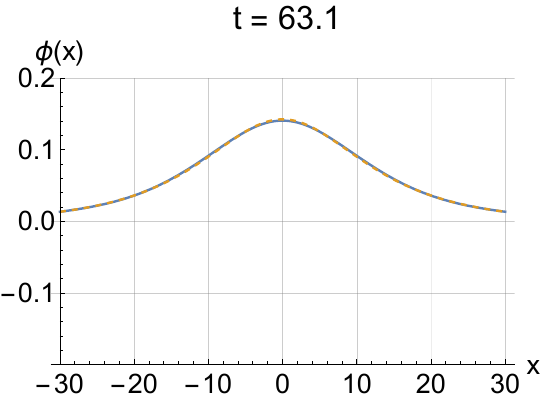}

\includegraphics[width=0.5\columnwidth]{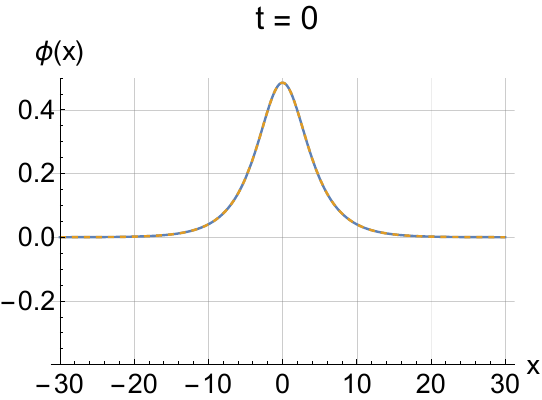}
\includegraphics[width=0.5\columnwidth]{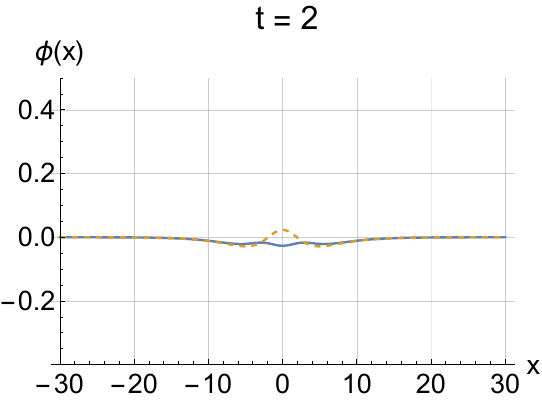}
\includegraphics[width=0.5\columnwidth]{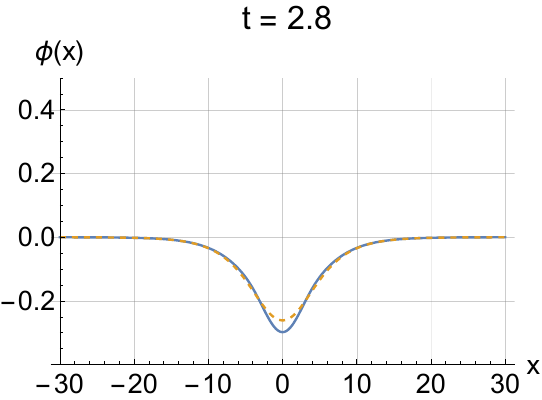}
\includegraphics[width=0.5\columnwidth]{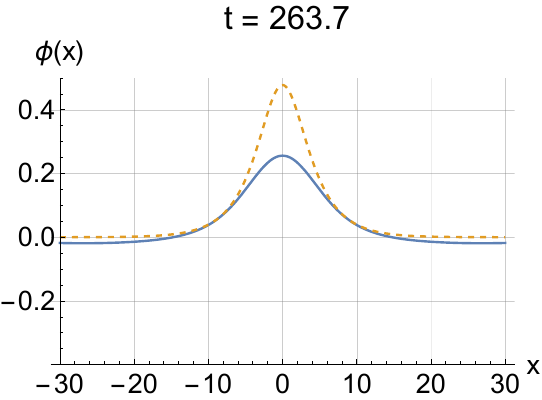}
\caption{\small Comparison between numerically found oscillon (blue) and renormalized solution (orange) for the single Q-ball solution in the double well $\phi^4$ theory. Upper: $\lambda=0.1$ and we plot the field profiles at $t=0, 2, 3.1$ and $t=63.1$.  Lower: $\lambda=0.3$ and $t=0, 2,2.8$ and $t=263.7$.}
\label{fig:dw-profile}
\end{center}
\end{figure*}

\begin{figure*}
\begin{center}
\includegraphics[width=1.0\columnwidth]{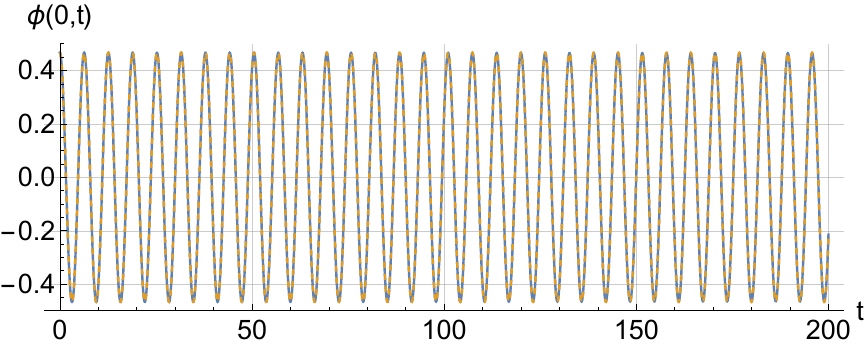}
\includegraphics[width=1.0\columnwidth]{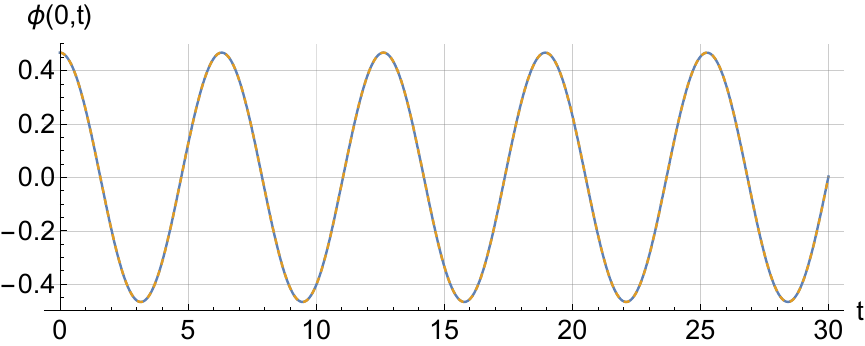}
\includegraphics[width=1.0\columnwidth]{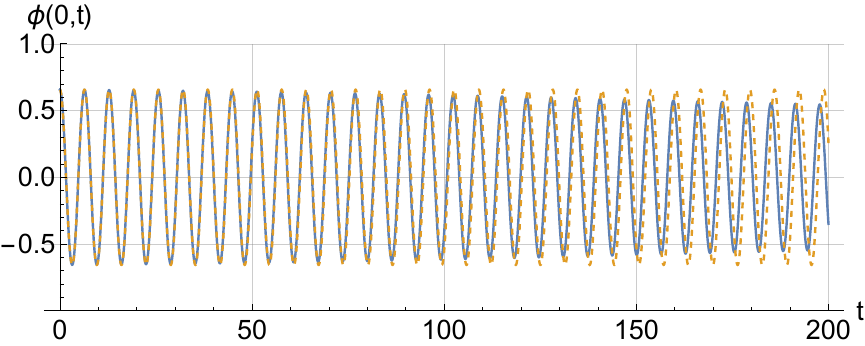}
\includegraphics[width=1.0\columnwidth]{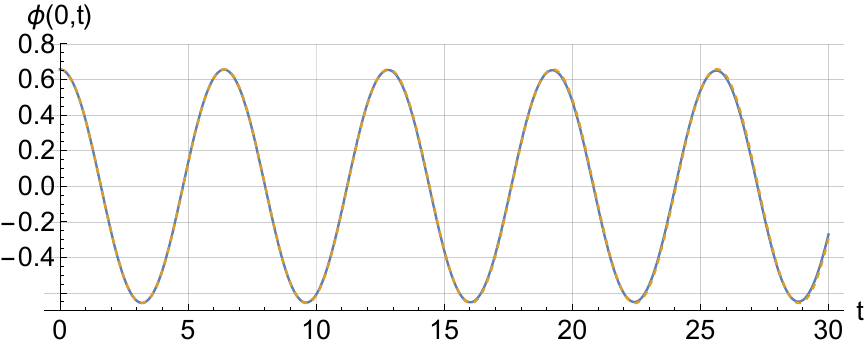}
\includegraphics[width=1.0\columnwidth]{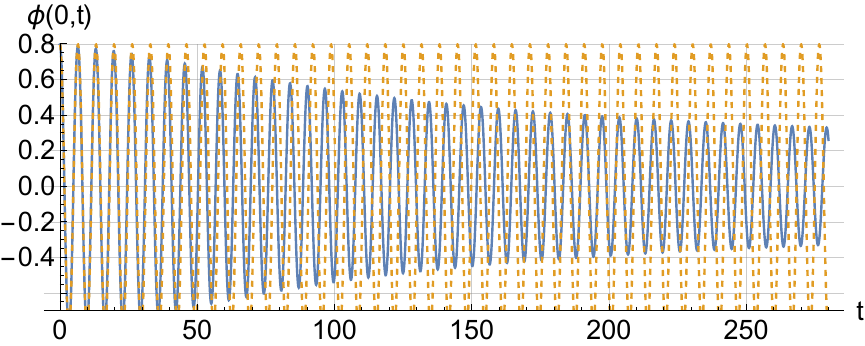}
\includegraphics[width=1.0\columnwidth]{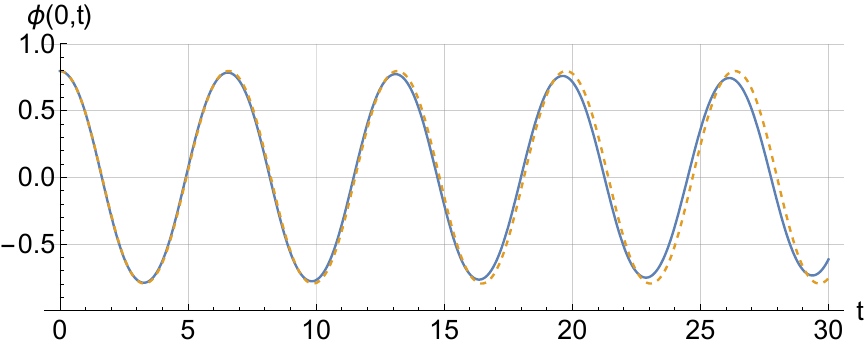}
\caption{\small Comparison between numerically found oscillon (blue) and renormalized solution (orange) for the single Q-ball solution in the exotic $\phi^6$ theory. We plot the value of the field at origin $\phi(x=0,t)$. Upper: $\lambda = 0.1$; Lower: $\lambda= 0.2$; Bottom: $\lambda= 0.3$. }
\label{fig:phi6}

\vspace*{0.5cm}

\includegraphics[width=0.5\columnwidth]{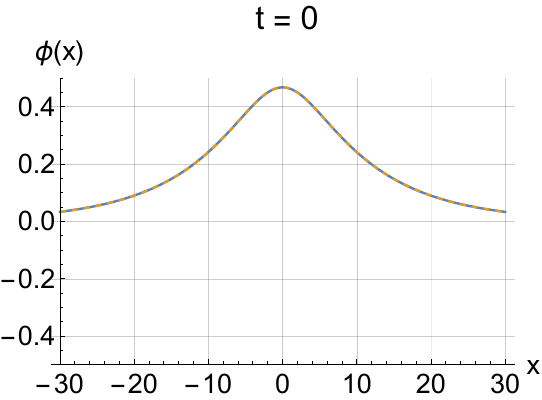}
\includegraphics[width=0.5\columnwidth]{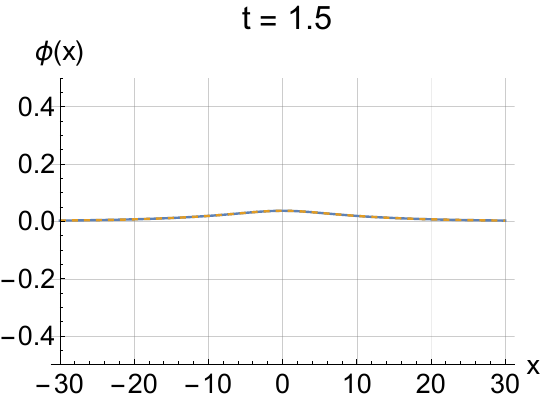}
\includegraphics[width=0.5\columnwidth]{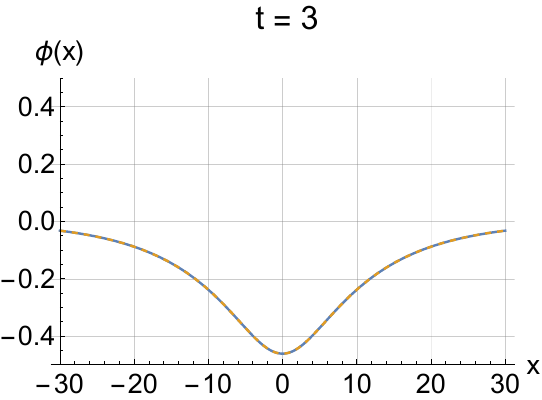}
\includegraphics[width=0.5\columnwidth]{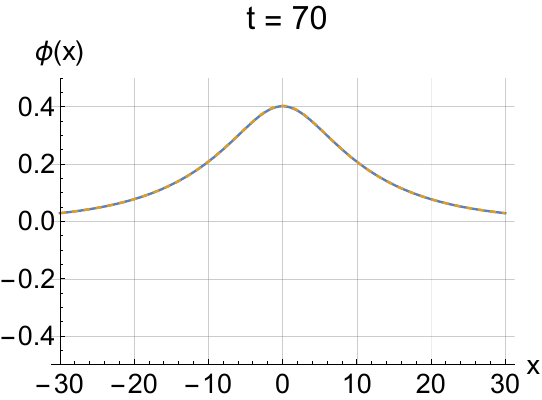}
\includegraphics[width=0.5\columnwidth]{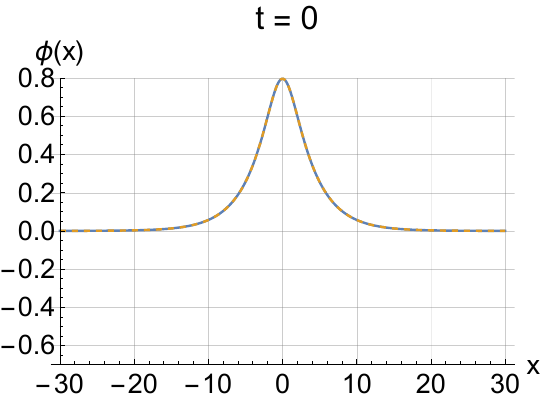}
\includegraphics[width=0.5\columnwidth]{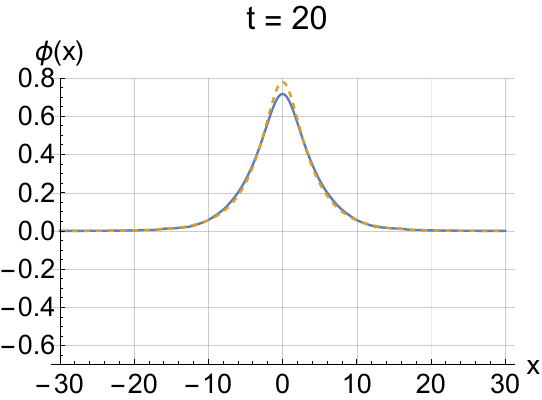}
\includegraphics[width=0.5\columnwidth]{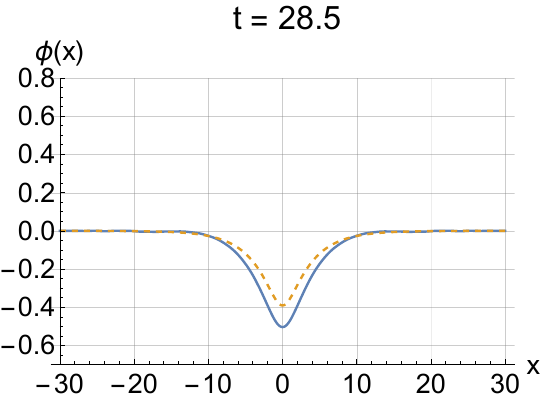}
\includegraphics[width=0.5\columnwidth]{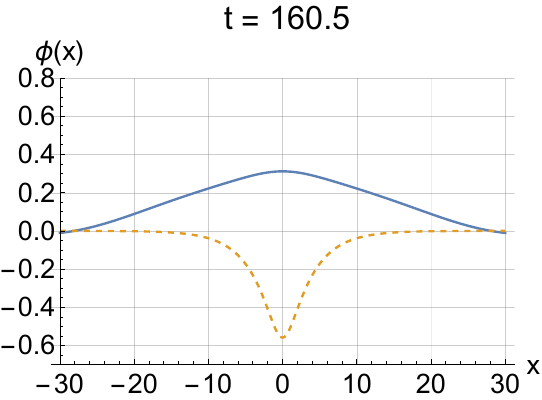}
\caption{\small Comparison between numerically found oscillon (blue) and renormalized solution (orange) for the single Q-ball solution in the exotic $\phi^6$ theory. Upper: $\lambda=0.1$ and we plot the field profiles at $t=0, 1.5, 3$ and $t=70$.  Lower: $\lambda=0.3$ and $t=0, 20,28.5$ and $t=160.5$.}
\label{fig:phi6-profile}
\end{center}
\end{figure*}

As the first example, we consider a scalar field theory with the potential $V(\phi) = \phi^2/2-\phi^3/3$, which was investigated in \cite{Manton:2023mdr} to expose a possible connection between spharelon solution and oscillons. This theory belongs to the generic class of the models with $a_3=1$ and all other terms vanishing.

The corresponding renormalized solution based on the single $Q$-ball (\ref{eq:qball}) reads
\begin{align}
\phi_{\rm R} = & \sqrt{\frac{12}{5}}\frac{\lambda \cos(\omega t)}{\cosh(\lambda x)} +\frac{2\lambda^2}{5}\frac{3-\cos(2\omega t)}{\cosh^2(\lambda x)}
\nonumber \\ \label{eq:renormalized1sol}
& +\sqrt{\frac{3}{5}}\frac{\lambda^3 \cos(3\omega t)}{10 \cosh^3(\lambda x)}\,.
\end{align}
This expression has the same terms as the approximate solution generated via FFHL expansion in \cite{Manton:2023mdr}. However, the solution generated by FFHL expansion has additional pieces:
\begin{equation}
\phi_{\rm Fodor} - \phi_{\rm R} = \sqrt{\frac{12}{5}}\frac{\lambda^3}{60}\frac{\cos(\omega t)}{\cosh(\lambda x)}\Bigl(94-\frac{119}{\cosh^2(\lambda x)}\Bigr)\,.
\end{equation}
Hence, although quite similar, in the unmodulated oscillon case, the RG approximation is not identical to the FFHL approximation, at least at the $\epsilon^3$ order. 

In Fig. \ref{fig:phi3} we plot $\phi(x=0,t)$ for the actual, numerically found oscillons and the renormalized solutions for different values of $\lambda$. As the initial configuration for the numerics, we took the renormalized solution  \refer{eq:renormalized1sol} at $t=0$. As we see, for small values of the initial amplitude we find a very good agreement. The numerical oscillon has rather negligible or very small modulations and therefore can be very well approximated by the configuration generated from the single $Q$-ball.

This is confirmed in Fig. \ref{fig:phi3-profile} where we plot the field profiles at certain times. Again, for small amplitudes, here $\lambda=0.2$, the actual numerical profiles agree very well with the renormalized configurations. The actual oscillon is very well approximated by our renormalized configuration based on the single $Q$-ball. We also see that the initial configuration does not relax to another oscillon. Thus there is very little radiation emitted during the initial time of the evolution. 

As the initial amplitude grows, and the oscillon becomes a large oscillon, the approximation is less and less accurate. The main reason is that the initial profile leads to a modulated oscillon which necessary requires two DoF. See, Fig. \ref{fig:phi3}  and Fig. \ref{fig:phi3-profile} for $\lambda=0.6$. Note that the modulated oscillon is not located only at the origin but has two centra. We also see that such initial conditions produce more radiation at the initial stage of the evolution. 

We also remark that the difference with the FFHL approximation is very small for the small oscillons. It is basically undistinguishable for the cases plotted in Fig. \ref{fig:phi3} and \ref{fig:phi3-profile}. As the amplitude grows the difference becomes to be visible but then we enter a regime where the true oscillon reveals the modulated structure. Hence, both the FFHL as well as the RG approximation based on the single $Q$-ball are no longer valid. 

\subsection{Inverse $\phi^4$ theory}

We obtain the same results for the inverse $\phi^4$ model $V(\phi) = \phi^2/2-\phi^4/4$ \cite{We}. Again, small oscillons oscillating around the first vacuum at $\phi=0$, are very well approximated by the renormalized solution based on the single $Q$-ball
\begin{align}
\phi_{\rm R} = & 2 \sqrt{\frac{2}{3}}\frac{ \lambda }{\cosh(\lambda x)} \cos(\omega t) - \frac{1}{6} \sqrt{\frac{2}{3} }  \frac{\lambda^3}{\cosh^3(\lambda x)} \cos (3\omega t)\,.
 \label{eq:renormalizedsol-4}
\end{align}
Larger oscillons reveal a modulated structure which cannot be captured by the single $Q$-ball, see Figs. \ref{fig:phi4} and \ref{fig:phi4-profile}.

\subsection{Double well $\phi^4$ theory}
The same picture repeats in the prototypical double well $\phi^4$ theory
$V(\phi) = \phi^2/2-\phi^3/2+\phi^4/4$. This form can be obtained for the usual expression $V=(1/8) (1-\tilde{\phi}^2)^2$, where the field $\tilde{\phi}$ is expanded around its vacuum value $\tilde{\phi}\equiv 1-\phi$.

Once again, the small, unmodulated oscillons are very well reproduced by the renormalized solution based on the single $Q$-ball solution of the complex $\phi^4$ equation, see Fig. \ref{fig:dw} and Fig. \ref{fig:dw-profile}. For increasing $\lambda$ the oscillons have more and more pronounced modulation structure which is not modeled by the simplest $Q$-ball state. 

We remark that in all these three examples the oscillons are qualitatively the same. They show very similar modulation patterns as the amplitude increases. From our point of view, this is an expected phenomenon as they emerge from the same RG $Q$-ball equation \refer{eq:RGrel}.

\subsection{Exotic $\phi^6$ theory}

Now we will analyze an exotic version of the $\phi^6$ potential, where the cubic as well as the quartic term is absent. For concreteness, we assume $V=\phi^2/2-\phi^6/6$. As we showed, this case leads to a different RG equation, namely Eq.~\refer{eq:rgn2}, that  has a non-linearity of the $\Psi |\Psi |^4$ type. 

The corresponding renormalized oscillon solution reads
\begin{align}
\phi_{\rm R} = &\, 2 \left( \frac{3}{10} \right)^{1/4} \left( \frac{ \lambda }{\cosh(2\lambda x)} \right)^{1/2} \cos(\omega t) \nonumber \\
& - \frac{5}{4}  \left( \frac{3}{10} \right)^{5/4} \left( \frac{ \lambda }{\cosh(2\lambda x)} \right)^{5/2}   \cos (3\omega t)\nonumber \\
& +  \frac{1}{12}  \left( \frac{3}{10} \right)^{5/4} \left( \frac{ \lambda }{\cosh(2\lambda x)} \right)^{5/2} \cos (5\omega t)\,,
 \label{eq:renormalizedsol-6}
\end{align}
with $\omega=\sqrt{1-\lambda^2}$. 

\begin{figure*}
\begin{center}
\includegraphics[width=2.0\columnwidth]{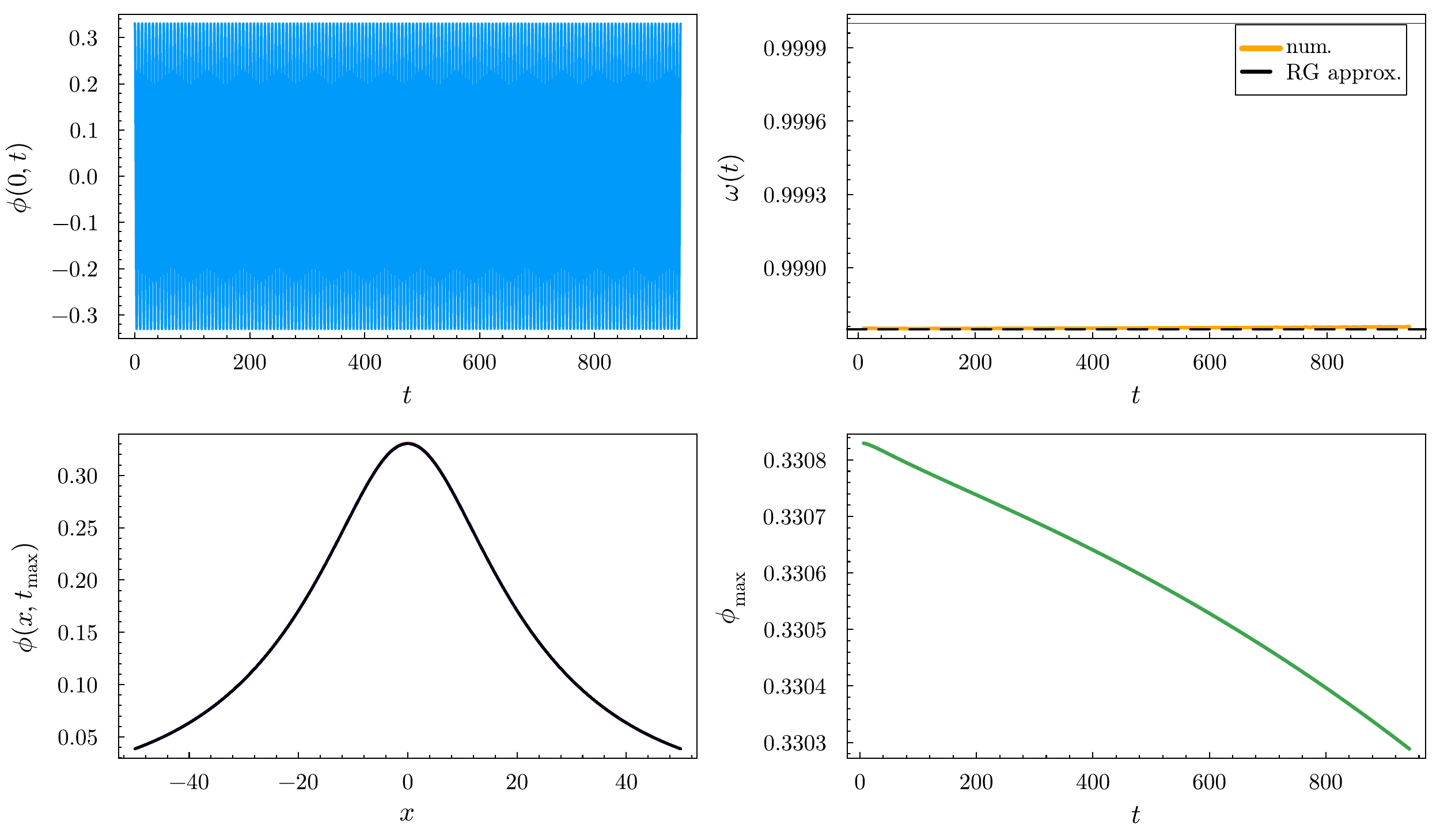}
\caption{\small A very long living oscillon in the exotic $\phi^6$ model. The initial condition is the renormalized solution $\phi_{\rm R}$ with $\lambda=0.05$.  Upper left: value fo the field at the origin. Upper right: measured frequency of oscillation (orange curve) vs the $Q$-ball frequency (black dashed line). Lower left: profiles of the field at different times. Lower right: evolution of the amplitude $\phi_{\rm max}$. }
\label{fig:exotic-osc}
\end{center}
\end{figure*}

First of all, it is not obvious that this model supports any oscillons. For example, it does not obey the condition \ref{eq:ourcond} and therefore, according to the FFHL approach, there should not exist any oscillons. Rather surprisingly, they do exist. We present some examples in Fig. \ref{fig:phi6} and  \ref{fig:phi6-profile}. Importantly, as predicted from our RG framework, they belong to a different universality class and differ a lot from the generic oscillons.  

In Fig. \ref{fig:exotic-osc} we present a very long living oscillon. It is obtained from the renormalized solution $\phi_{\rm R}$ based on the single $Q$-ball with $\lambda=0.05$. The measured frequency of the oscillations is clearly below the mass threshold and very well agrees with the $Q$-ball frequency. Also the profile is well localized and very well preserved during the almost periodic time evolution. The amplitude of the oscillations decreases in an extremely slow manner.  Definitely, it is a well behaving oscillon. It is also very well approximated by the renormalized solution. 

The most striking feature is the lack of amplitude modulation while the initial amplitude grows. On the contrary, for different values of $\lambda$, the generated oscillon all the time has only one fundamental frequency but it becomes less and less stable for bigger $\lambda$. This means that after some time, which is shorter for increasing $\lambda$, the oscillon smoothly decays.  Its amplitude decreases and its width grows which is an effect of loosing energy via radiation, see Fig. \ref{fig:phi6-profile}. 

To conclude, undoubtedly these oscillons present different qualitative features than the generic ones. This can be easily explained within our RG scheme, where they originate from a different $Q$-ball equation. They simply belong to a different universality class.

\section{Amplitude modulation and Integrable theory of $Q$-balls}
\subsection{Integrable complex sine-Gordon}
\begin{figure*}
\begin{center}
\includegraphics[width=1.0\columnwidth]{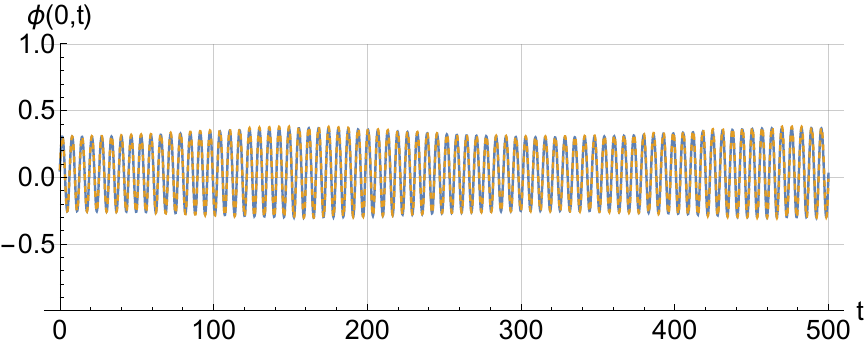}
\includegraphics[width=1.0\columnwidth]{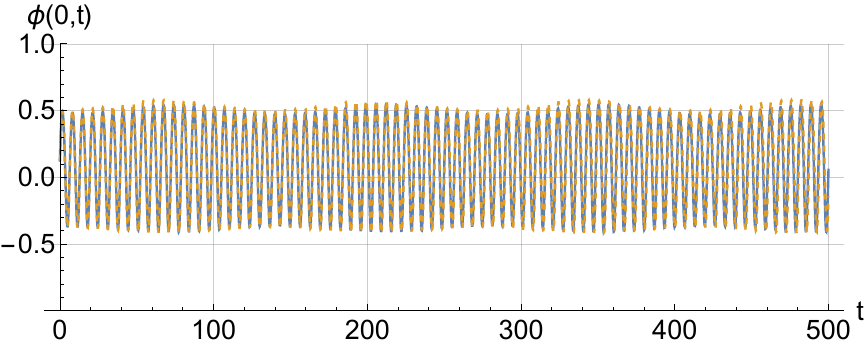}
\includegraphics[width=1.0\columnwidth]{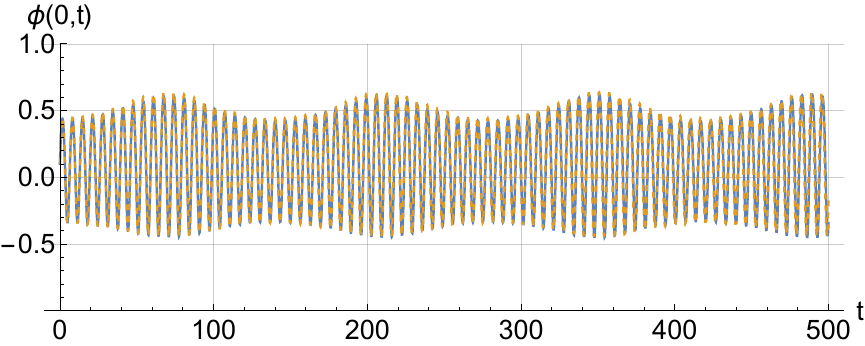}
\includegraphics[width=1.0\columnwidth]{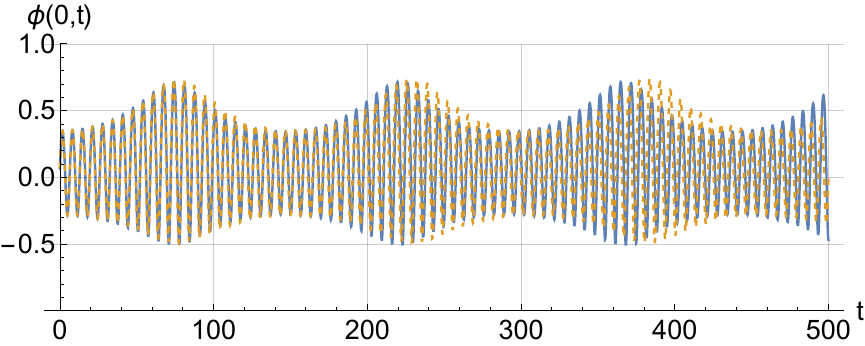}
\includegraphics[width=1.0\columnwidth]{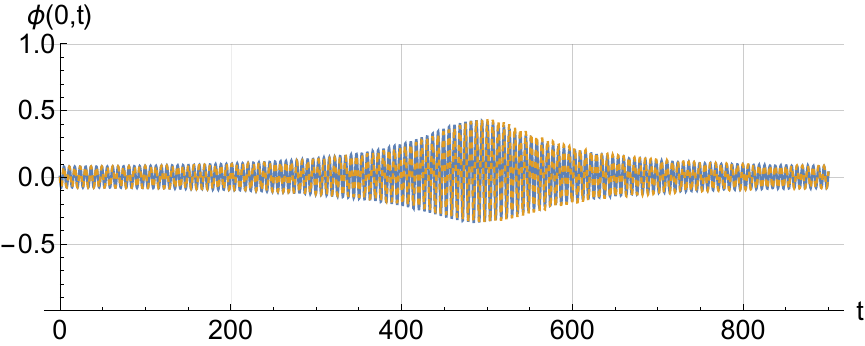}
\includegraphics[width=1.0\columnwidth]{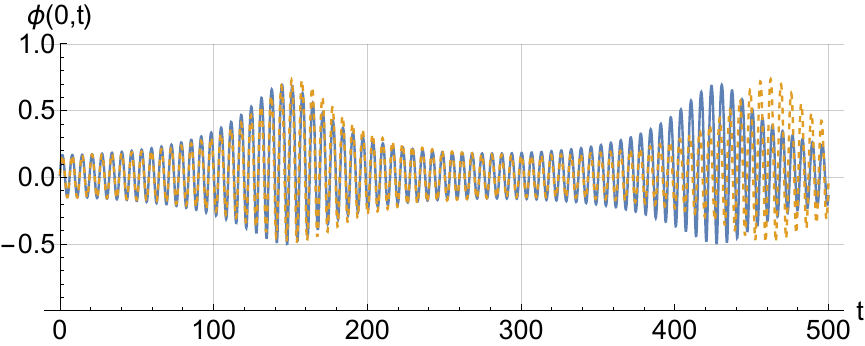}
\caption{\small Comparison between numerically found modulated oscillon (blue) and renormalized solution (orange) for the two Q-ball solution in the $\phi^3$ theory. We plot value of the field at origin $\phi(x=0,t)$. Upper left: $\lambda_1 = 0.2, \lambda_2=-0.02$; Upper right: $\lambda_1 = 0.3, \lambda_2=-0.02$; Lower left: $\lambda_1 = 0.3, \lambda_2=-0.05$; Lower right: $\lambda_1 = 0.3, \lambda_2=-0.1$; Bottom left: $\lambda_1 = 0.1, \lambda_2=-0.15$; Bottom right: $\lambda_1 = 0.25, \lambda_2=-0.15$. }
\label{fig:two}

\vspace*{0.5cm}

\includegraphics[width=0.51\columnwidth]{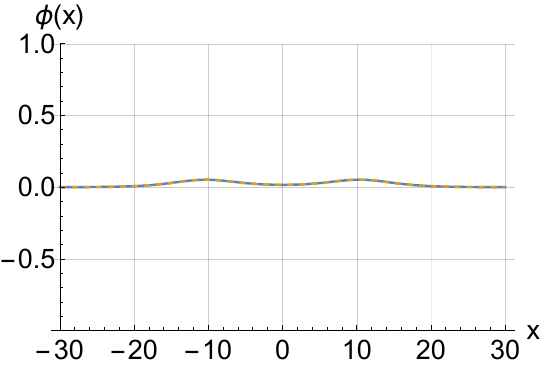}
\includegraphics[width=0.51\columnwidth]{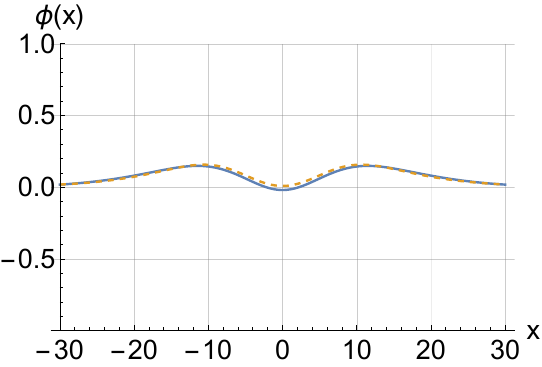}
\includegraphics[width=0.51\columnwidth]{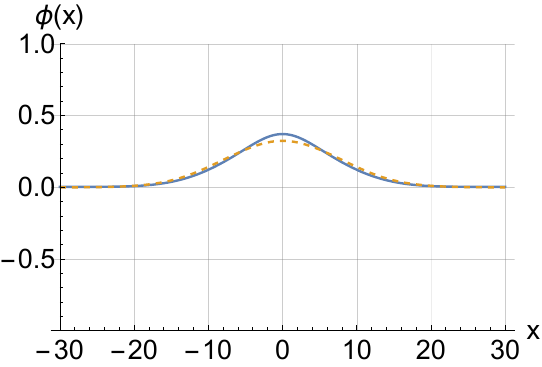}
\includegraphics[width=0.51\columnwidth]{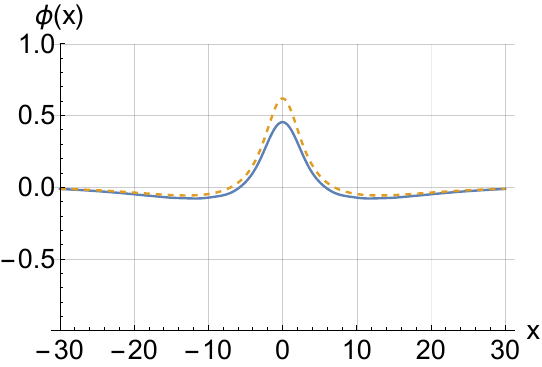}
\includegraphics[width=0.5\columnwidth]{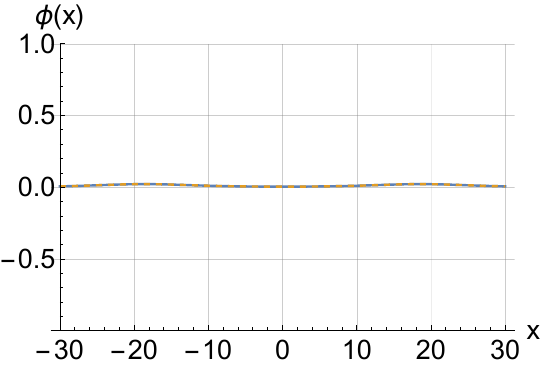}
\includegraphics[width=0.5\columnwidth]{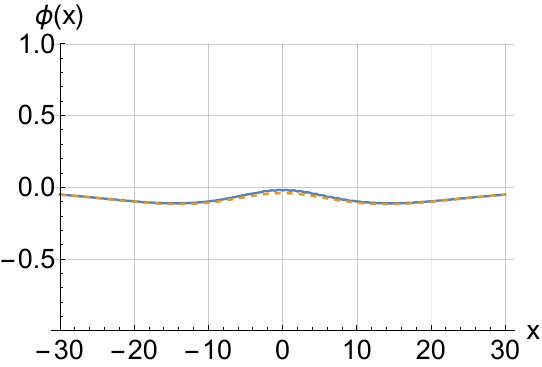}
\includegraphics[width=0.5\columnwidth]{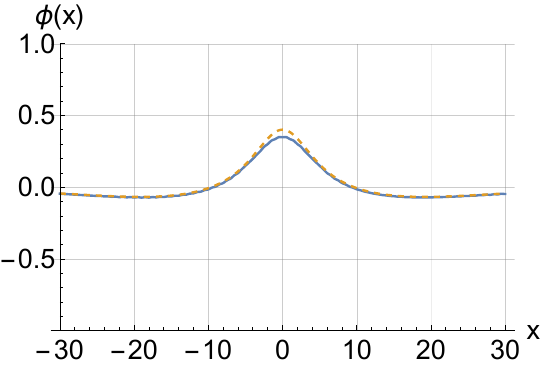}
\includegraphics[width=0.5\columnwidth]{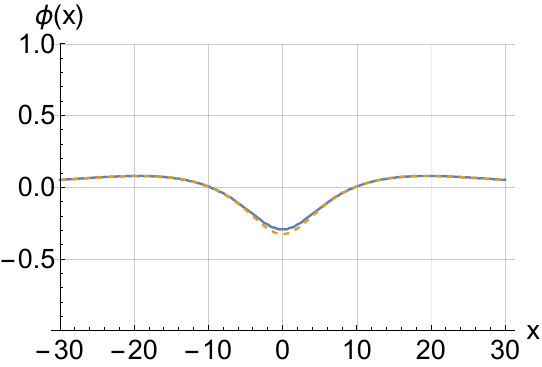}

\caption{\small Comparison between numerically found modulated oscillon (blue) and the renormalized solution (orange) for the two Q-ball solution in the $\phi^3$ theory. Upper:$\lambda_1 = 0.25, \lambda_2=-0.15$ and we plot the field profiles at $t=0, 103, 105$ and $t=150$.  Lower: $\lambda_1 = 0.1, \lambda_2=-0.15$ and $t=0, 400,500$ and $t=510$.}
\label{fig:two-profile}
\end{center}
\end{figure*}

\begin{figure*}
\begin{center}
\includegraphics[width=1.8\columnwidth]{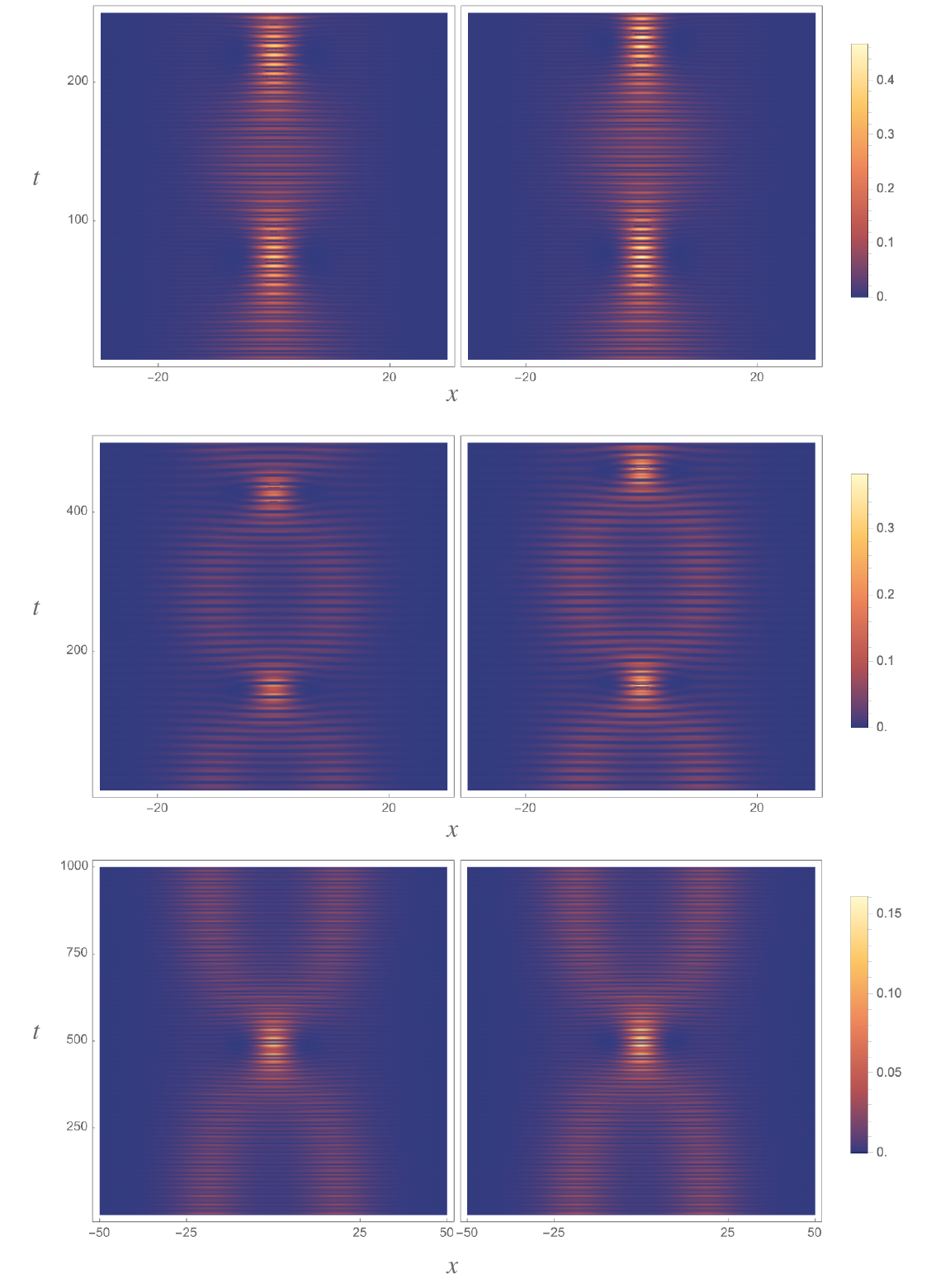}
\caption{\small Comparison between numerically found modulated oscillon (left) and the renormalized solution obtained from the two Q-ball solution (right) in the $\phi^3$ theory. We plot the absolute value of $(\partial^2+1)\phi$ as a function of $x$ and $t$. Upper: $\lambda_1 = 0.3, \lambda_2=-0.1$; Lower: $\lambda_1 = 0.25, \lambda_2=-0.15$; Bottom: $\lambda_1 = 0.1, \lambda_2=-0.15$. }
\label{fig:den}
\end{center}
\end{figure*}

\begin{figure*}
\begin{center}
\includegraphics[width=1.0\columnwidth]{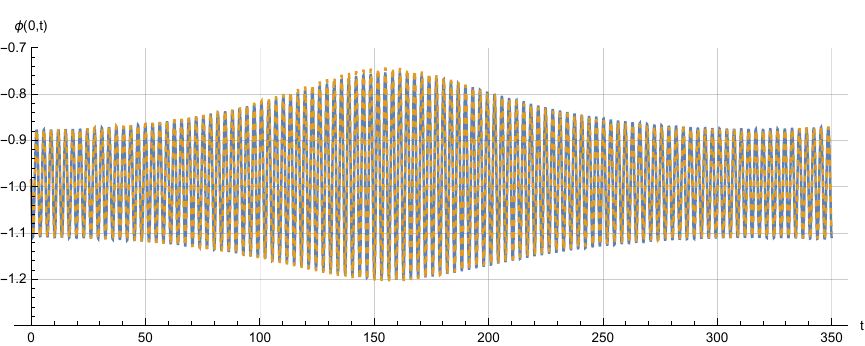}
\includegraphics[width=1.0\columnwidth]{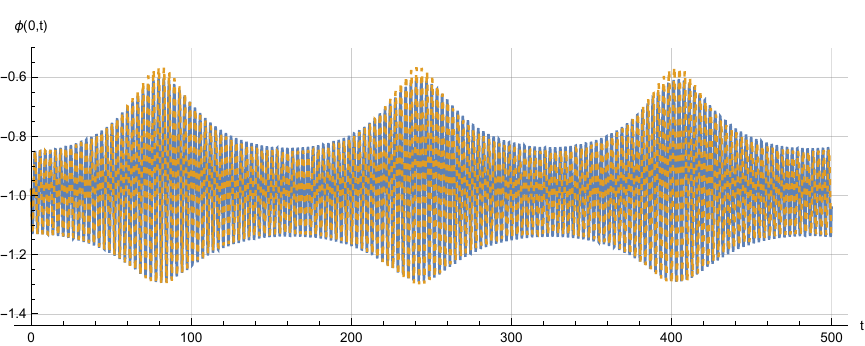}
\includegraphics[width=1.0\columnwidth]{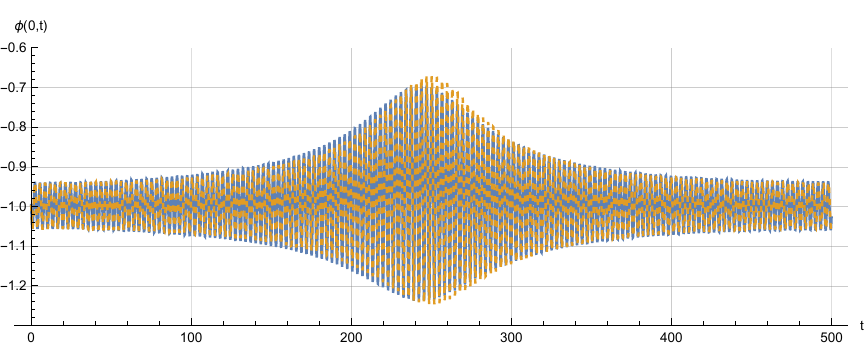}
\includegraphics[width=1.0\columnwidth]{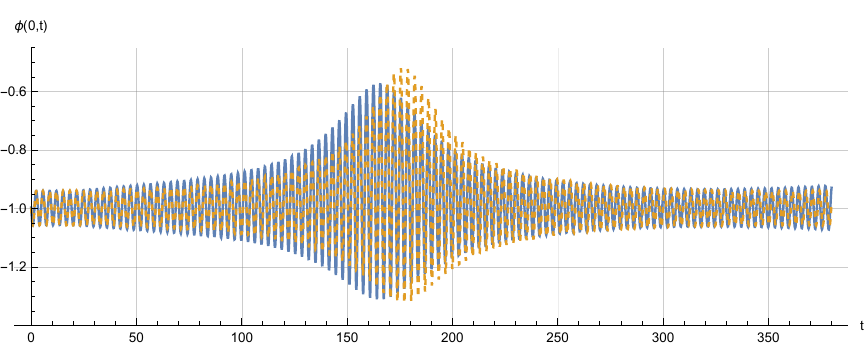}

\caption{\small Comparison between numerically found modulated oscillon (blue) and renormalized solution (orange) for the two Q-ball solution in the duble well $\phi^4$ theory. We plot value of the field at origin $\phi(x=0,t)$. Upper left: $\lambda_1 = 0.05, \lambda_2=-0.15$; Upper right: $\lambda_1 = 0.1, \lambda_2=-0.22$; Lower left: $\lambda_1 = 0.1, \lambda_2=-0.15$; Lower xright: $\lambda_1 = 0.15, \lambda_2=-0.2$. }
\label{fig:profile-dw}

\vspace*{0.5cm}

\includegraphics[width=1.0\columnwidth]{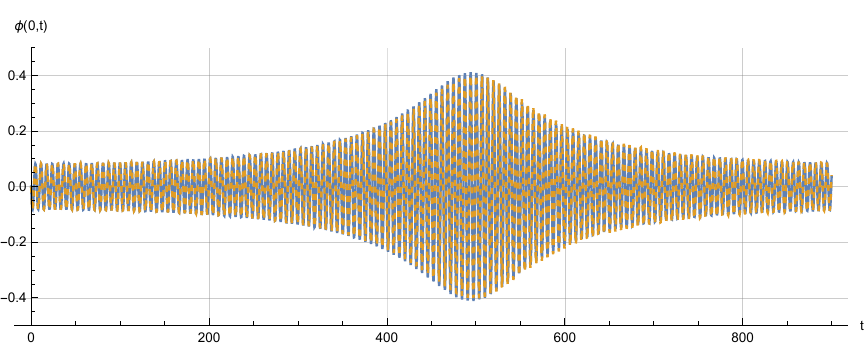}
\includegraphics[width=1.0\columnwidth]{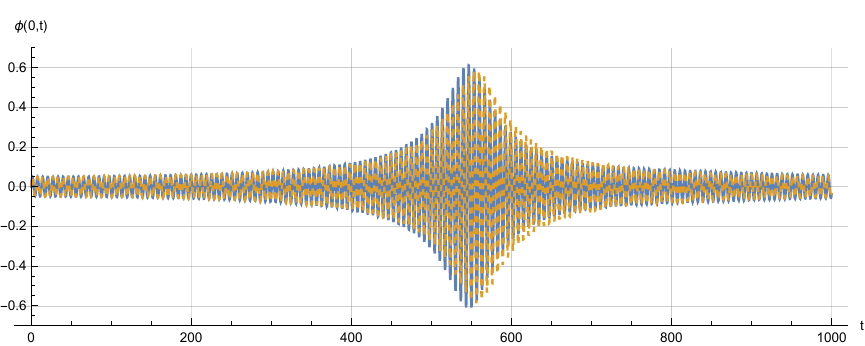}
\includegraphics[width=1.0\columnwidth]{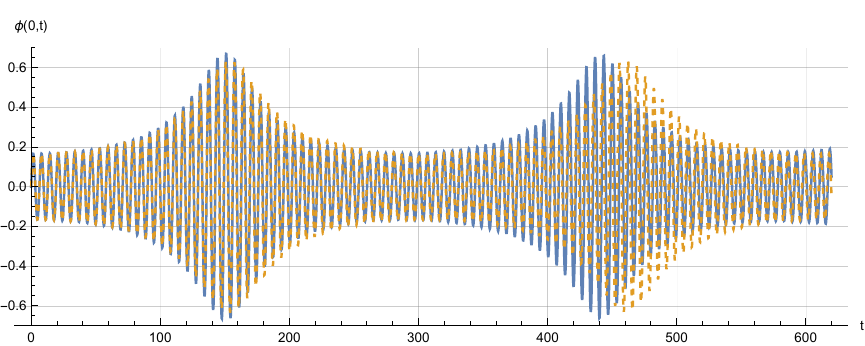}
\includegraphics[width=1.0\columnwidth]{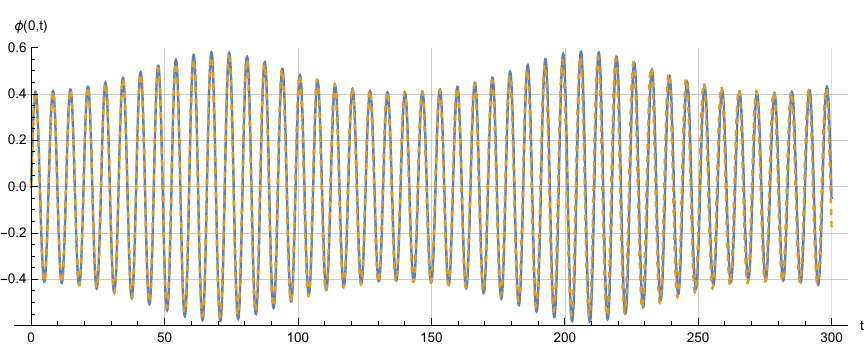}

\caption{\small Comparison between numerically found modulated oscillon (blue) and renormalized solution (orange) for the two Q-ball solution in the reverse $\phi^4$ theory. We plot value of the field at origin $\phi(x=0,t)$. Upper left: $\lambda_1 = 0.1, \lambda_2=-0.15$; Upper right: $\lambda_1 = 0.2, \lambda_2=-0.17$; Lower left: $\lambda_1 = 0.25, \lambda_2=-0.15$; Lower right: $\lambda_1 = 0.3, \lambda_2=-0.05$. }
\label{fig:profile-inv}

\end{center}
\end{figure*}

\begin{figure*}
\begin{center}
\includegraphics[width=1.5\columnwidth]{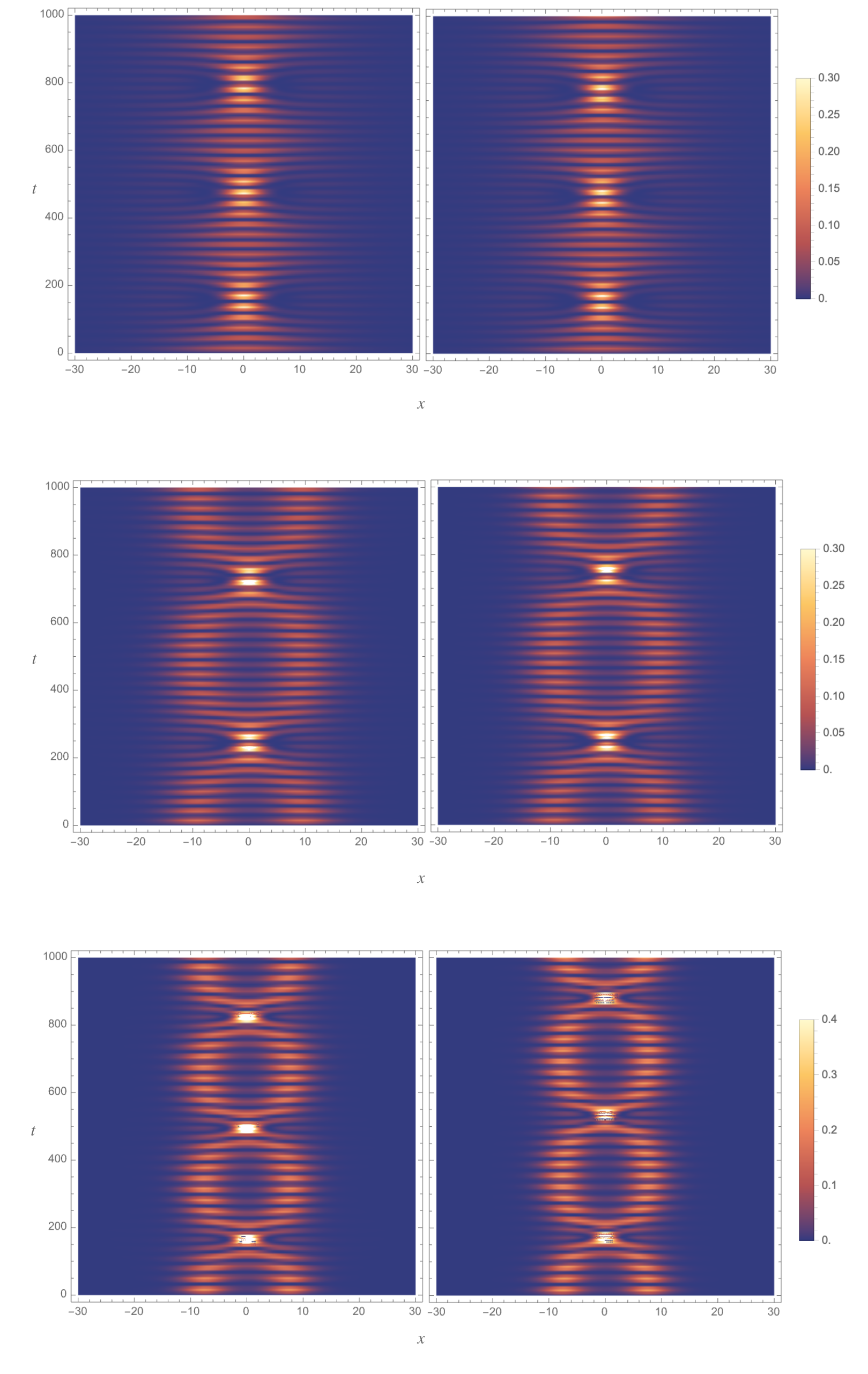}
\caption{\small Comparison between numerically found modulated oscillon (left) and the renormalized solution obtained from the two Q-ball solution (right) in the double well $\phi^4$ theory. We plot the absolute value of $(\partial^2+1)(\phi+1)$ as a function of $x$ and $t$. Upper: $\lambda_1 = 0.05, \lambda_2=-0.15$; Lower: $\lambda_1 = 0.1, \lambda_2=-0.15$; Bottom: $\lambda_1 = 0.15, \lambda_2=-0.2$. }
\label{fig:dw-den}
\end{center}
\end{figure*}

\begin{figure*}
\begin{center}
\includegraphics[width=1.5\columnwidth]{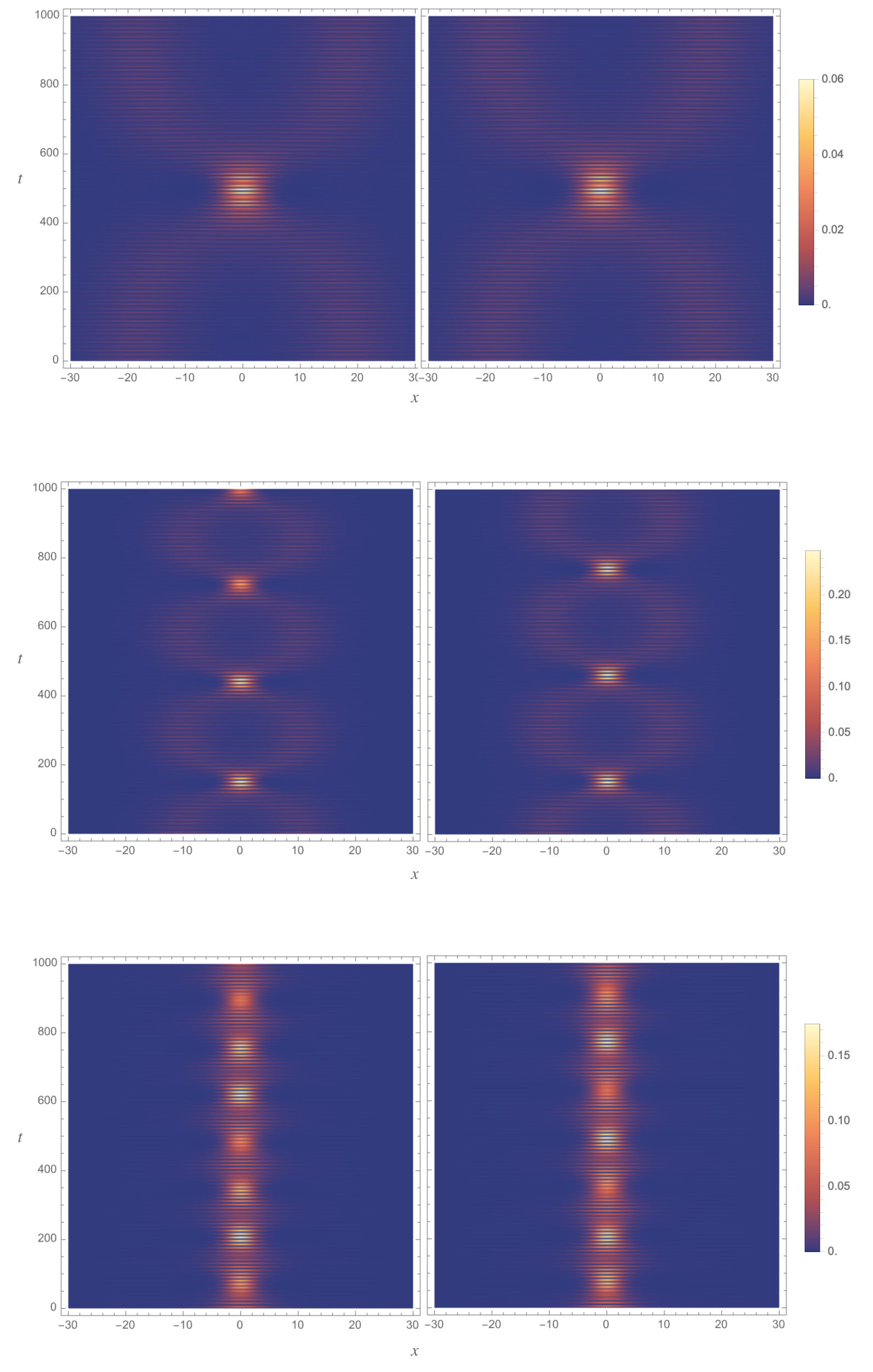}

\caption{\small Comparison between numerically found modulated oscillon (left) and the renormalized solution obtained from the two Q-ball solution (right) in the inverse $\phi^4$ theory. We plot the absolute value of $(\partial^2+1)\phi$ as a function of $x$ and $t$. Upper: $\lambda_1 = 0.1, \lambda_2=-0.15$; Lower: $\lambda_1 = 0.25, \lambda_2=-0.15$; Bottom: $\lambda_1 = 0.3, \lambda_2=-0.05$. }
\label{fig:inv-den}
\end{center}
\end{figure*}

It is a matter of fact that, for the case of the generic potentials, the universal RG $Q$-ball equation can be approximated by the integrable complex sine-Gordon ($\mathbb{C}$sG) equation:
\begin{equation}
\bigl(\partial^2 +1\bigr)\Psi = \Psi |\Psi|^2 - \bar\Psi \frac{\partial_\mu \Psi \partial^\mu \Psi}{1- |\Psi|^2}\,.
\end{equation} 
The sense, in which this equation is ``close'' to RG equation \refer{eq:RGrel} is that it has exactly the same $Q$-ball solution given in Eq.~\refer{eq:qball}. Furthermore, the Lagrangian describing $\mathbb{C}$sG model differs from \refer{eq:lag1} by a multiplicative factor of the form $1+ \mathcal{O}\bigl(|\Psi|^2\bigr)$, i.e.,
\begin{equation}
\mathcal{L}_{\mbox{$\mathbb{C}$sG}} = \frac{1}{1-|\Psi|^2}\Bigl(\partial_\mu \bar\Psi \partial^\mu \Psi - |\Psi |^2+|\Psi |^4\Bigr)\,.
\end{equation}

Due to its integrability, we can write down multi-soliton solutions in a closed form. The simplest example is the two $Q$-ball solution  \cite{Bowcock:2008dn}
\begin{widetext}
\begin{equation}
\Psi_{12} = \frac{(\Psi_1 \delta_1-\Psi_2 \delta_2)(\bar\Psi_{1}^{K} \delta_2-\bar\Psi_2^{K} \delta_1)-(\Psi_2 \delta_1-\Psi_1\delta_2)(\Psi_2^{K}\delta_2-\Psi_1^K \delta_1)}{\delta_1^2+\delta_2^2-\delta_1\delta_2 \bigl(\Psi_1 \bar\Psi_2+\bar\Psi_1\Psi_2+\Psi_1^K\bar\Psi_2^K+\bar\Psi_1^K\Psi_2^K\bigr)}\,,
\end{equation}
\end{widetext}
where $\Psi_{1,2}$ is a single Q-ball solution with an arbitrary scale parameter $\lambda_{1,2}$, boost $v_{1,2}$ and position $a_{1,2}$, while 
\begin{equation}
\delta_{1,2} \equiv \sqrt{\frac{1-v_{1,2}}{1+v_{1,2}}}.
\end{equation}

Furthermore, the $\Psi_{1,2}^K$'s are the associated kink solutions, given as
\begin{multline}
\Psi_{1,2}^{K} \equiv -\lambda_{1,2}\tanh\biggl(\frac{\lambda_{1,2}}{\sqrt{1-v_{1,2}^2}}\bigl(x- v_{1,2} t-a_{1,2}\bigr)\biggr)\\
-\I \sqrt{1-\lambda_{1,2}^2}\,.
\end{multline}

For our purposes and for simplicity, let us consider the coincident, stationary two Q-ball solutions, i.e., $v_{1,2} = a_{1,2} = 0$, describing a bound state of two $Q$-balls located on top of each other and centered at the origin with zero total momentum.

This solution can be explicitly given in the following compact form 
\begin{widetext}
\begin{equation}
\Psi_{12} = \frac{\I \bigl(\omega_1 -\omega_2\bigr)\Bigl(\frac{\lambda_1}{\cosh(\lambda_1 x)}\e^{\I \omega_1 t}-\frac{\lambda_2}{\cosh(\lambda_2 x)}\e^{\I \omega_2 t}\Bigr)}
{1-\omega_1\omega_2 -\lambda_1\lambda_2 \Bigl(\tanh(\lambda_1 x) \tanh(\lambda_2 x)+\frac{\cos(t(\omega_1-\omega_2))}{\cosh(\lambda_1 x)\cosh(\lambda_2 x)}\Bigr)}\,, \;\;\;  \omega_{1,2}\equiv \sqrt{1-\lambda_{1,2}^2}.
\end{equation}
\end{widetext}

Let us now treat this two-$Q$-ball state as a seed for the renormalized solution. 

Of course, this makes sense only in the case of the models belonging to the universality class belonging to the complex $|\Psi|^4$ model. For the exotic $\phi^6$ potential which leads to the RG $Q$-ball theory with $|\Psi|^6$ and without $|\Psi|^4$ term, the connection with the $\mathbb{C}$sG and the integrability is lost. This clearly indicate a deep, qualitative difference between oscillons belonging to these two universality classes. As we have already seen, this is indeed the case. 

\subsection{Modulations in $\phi^3$ oscillons}

We consider the simplest model belonging to the generic universality class, i.e., $\phi^3$ theory. As we previously verified the single $Q$-ball solution was not able to capture the modulated oscillons. Now, we use the two $Q$-ball state and insert it into the renormalized solution
 \begin{equation}\label{eq:2Qrenorm}
\phi_{\rm R}  =
\sqrt{\frac{3}{5}}\Psi_{12} -\frac{1}{2}\Psi_{12}^2+\frac{3}{5}|\Psi_{12}|^2+\frac{1}{20}\sqrt{\frac{3}{5}}\Psi_{12}^3 + \cc
\end{equation}
Again we use $\phi_{\rm R}$ as the initial data for the numerical computations. The results are impressive.

First of all the two $Q$-ball renormalized solution with great accuracy reproduces the amplitude modulations for the wide range of oscillons, see Fig. \ref{fig:two} and \ref{fig:two-profile}. The match is spectacular for small and moderate modulations, that is for the case where $\lambda_2$ is close to 0. It is also good for very large modulations, see Fig. \ref{fig:two} bottom panels. Undoubtedly, the modulation is an effect of a motion of two unmodulated oscillons, each arising from its $Q$-ball. 

This remarkable agreement concerns not only the oscillation of the value of the field at the origin, Fig. \ref{fig:two}, but also the entire field profiles, Fig. \ref{fig:two-profile} and Fig. \ref{fig:den} (concretely, we plot the absolute value of $(\partial^2+1)\phi$ which shows the field evolution in the  clearest way). Here, the inner, double centered structure of the modulated oscillon is very well visible and very well reproduced by our renormalized solution. 

Importantly, the initial conditions provide clear oscillons. There is very little radiation at the beginning of the evolution. This is another sign that our approximated renormalized solution is very close to the true one.  

\subsection{Further examples}
Finally we consider modulated oscillons in the double well and reverse $\phi^4$ model.  Again, modulated oscillons can be very well approximated by the renormalized solution based on the two $Q$-ball state, see Fig.  \ref{fig:profile-dw}, Fig. \ref{fig:profile-dw} where we plot the value of the field at the origin and Fig. \ref{fig:dw-den}, \ref{fig:inv-den} where we compare the evolution of the absolute value of $(\partial^2+1)\phi$ in the full space. 

Of course, in the case of the exotic $\phi^6$ model, there is no reason to use the doble $Q$-solution as the seed for the oscillons. The RG derived $Q$-ball equation has rather nothing to do with the $\mathbb{C}$sG model. This explains the nonexistence of modulated oscillons in this model. 

\section{Corrections to RG perturbation expansions}
\label{sec:corr}

In this section, we shall look at the higher-order corrections to oscillons that are produced when we carry out the RGPE algorithm to the higher powers in the perturbation parameter. Namely, we show that, at the next leading order, the RG equations pick up generically a $\Psi |\Psi|^4$ term whose coefficient depends on the model, signaling departure from both universality and integrability. Of course, these corrections are sub-leading and the observed departure from the universality is a secondary effect. 

As a result, the $Q$-ball solution obtains model-dependent corrections. In the case of sine-Gordon model, we verify that these corrections match precisely with the expansion of the exact breather solution. This is a very important consistency check of the RGPE framework. 

\subsection{RGPE to $\varepsilon^5$ order.} 

Let us consider a generic potential with a vacuum at $\phi = 0$, which can be always enforced by shifting the field.
The part of the equation of motion relevant to the 5-th order expansion reads
\begin{equation}
\bigl(\partial^2+1\bigr)\phi = a_3 \phi^2 + a_4 \phi^3 +a_5 \phi^4 +a_6 \phi^5\,,
\end{equation}
where we, also without loss of generality, rescaled coordinates so that the perturbative mass is $m^2 =1$.

Plugging in the expansion
\begin{equation}
\phi = \varepsilon \phi_1+\varepsilon^2 \phi_2 + \varepsilon^3 \phi_3 + \varepsilon^4 \phi_4+\varepsilon^5 \phi_5 +\ldots
\end{equation}
we solve the above equation of motion order by order up to $\varepsilon^5$.

The bare expansion coefficients are found to be 
\begin{align}
\phi_1 & = A_0 \e^{\I \theta} + \cc\,, \\
\phi_2 & = \alpha_2 A_0^2 \e^{2\I \theta}+\beta_2 |A_0|^2/2 +\cc\,, \\
\phi_3 & = \alpha_3 A_0^3 \e^{3\I \theta}+ A_0 |A_0|^2 \e^{\I \theta}\mathcal{S}_1 + \cc\,, \\
\phi_4 & = \alpha_4 A_0^4 \e^{4\I\theta} +A_0^2|A_0|^2 \e^{2\I\theta}\mathcal{S}_2 \nonumber \\
&\phantom{=} + \frac{1}{2}|A_0|^4 \mathcal{S}_3+\cc\,, \\
\phi_5 & = \alpha_5 A_0^5 \e^{5\I \theta} + A_0^3 |A_0|^2 \e^{3\I \theta} \mathcal{S}_4 \nonumber \\
& \phantom{=}+ A_0 |A_0|^4 \e^{\I \theta} \mathcal{S}_5 +\cc\,,
\end{align}
where 
{\small \begin{gather}
\alpha_2 = -\frac{a_3}{3}\,, \quad \beta_2 = 2a_3\,, \quad \alpha_3 = \frac{2a_3^2-3a_4}{24}\,, \\
\alpha_4 = \frac{-10 a_3^3+45 a_3 a_4 -36 a_5}{540}\,, \\
\alpha_5 = \frac{5a_3^4}{1296}-\frac{5a_3^2a_4}{144}+\frac{a_4^2}{64}+\frac{11a_3a_5}{180}-\frac{a_6}{24}\,,
\end{gather}}
and where the secular terms obey the formulae:
{\small \begin{align}
2\I \partial_\theta \mathcal{S}_1 + \partial^2 \mathcal{S}_1 & = \frac{10 a_3^2 +9a_4}{3}\,, \\
4\I \partial_\theta \mathcal{S}_2 +\partial^2 \mathcal{S}_2 & = 3 \mathcal{S}_2 +2a_3 \mathcal{S}_1 +4a_5 +\frac{15a_3a_4}{4}-\frac{7a_3^3}{6}\,, \\
 \partial^2 \mathcal{S}_3+\mathcal{S}_3 & = 2a_3\bigl(\mathcal{S}_1+\bar{\mathcal{S}}_1\bigr)+6a_5+10a_3a_4+\frac{38a_3^3}{9}\,, \\
6\I \partial_\theta \mathcal{S}_4 + \partial^2 \mathcal{S}_4 & = 8 \mathcal{S}_4  +2a_3\mathcal{S}_2+\Bigl(3a_4-\frac{2a_3^2}{3}\Bigr)\mathcal{S}_1+5a_6\nonumber \\
& \phantom{=}+\frac{58a_3a_5}{15}-\frac{3a_4^2}{4}-\frac{7a_3^2a_4}{2}+\frac{8a_3^4}{27}\,, \\
2\I \partial_\theta \mathcal{S}_5 + \partial^2 \mathcal{S}_5 & = 2a_3\bigl(\mathcal{S}_3+\mathcal{S}_2\bigr)+\Bigl(3a_4-\frac{2a_3^2}{3}\Bigr)\bar{\mathcal{S}}_1\nonumber \\ 
 & \phantom{=}+\bigl(4a_3^2+6a_4\bigr)\mathcal{S}_1 +10 a_6 +\frac{56a_3a_5}{3}
\nonumber \\
&\phantom{=} -\frac{3a_4^2}{8}+9a_3^2a_4-\frac{a_3^4}{18}\,.
\end{align}}

Now we redefine the amplitude as
\begin{align}\label{eq:dresseda}
A_0 & = A \Bigl(1-\varepsilon^2 |A|^2\mathcal{S}_1^0+ \varepsilon^4 |A|^4  \mathcal{X}\Bigr)\,,
\end{align}
where $A \equiv A(\theta_0, \bar\theta_0)$ and where we denoted:
\begin{align}
\mathcal{X} & = -\mathcal{S}_5^{0}+2\bigl(\mathcal{S}_1^{0}\bigr)^2+|\mathcal{S}_1^0|^2+\mathcal{Z}\,.
\end{align}
Here, $\mathcal{Z}$ is, so far, an unknown function of renormalized scales $\theta_0$ and $\bar\theta_0$.

The strategy is now slightly different. We use the freedom in defining the dressed amplitude not for (fully) removing the secular dependence of the renormalized solution, but to make the minimal  subtraction of unwanted secular terms from the RG equation.  This is because, in a sense, we are not calculating the corrections to the renormalized solution itself, but only to the RG equation.

Indeed, the renormalized solution -- which we obtain by setting all $\mathcal{S}_i^0 = \mathcal{S}_i$ and $\theta_0 = \theta$, $\bar\theta_0 = \bar\theta$ --  has the form
\begin{align}
\phi_{\rm R} & = \varepsilon A \e^{\I \theta}+\alpha_2 \varepsilon^2 A^2 \e^{2\I \theta}+ \beta_2 |A|^2/2 +\alpha_3 \varepsilon^3 A^3 \e^{3\I \theta}+ \cc \nonumber \\
& \phantom{=} + \alpha_4 \varepsilon^4 A^4\e^{4\I \theta} +\varepsilon^4 A^2|A|^2\e^{2\I \theta} \bigl(2\alpha_2 \mathcal{S}_1-\mathcal{S}_2\bigr) +\cc \nonumber \\
&\phantom{=}+
\varepsilon^4|A|^4\bigl(\mathcal{S}_3-\beta_2 \mathcal{S}_1-\beta_2 \bar{\mathcal{S}_1}\bigr)/2 + \alpha_5 \varepsilon^5 A^5\e^{5\I\theta}+\cc
\nonumber \\
&\phantom{=} +\varepsilon^5 A^3|A|^2 \e^{3\I\theta}\bigl(\mathcal{S}_4-3\alpha_3 \mathcal{S}_1\bigr) + \varepsilon^5 A|A|^4 \e^{\I\theta} \mathcal{Z} +\cc
\end{align}
Note that the residual secular dependence makes this solution sensible only up to the $\varepsilon^3$ terms. Therefore, for practical purposes, we should truncate the above expression at $\varepsilon^3$ (or keep only those higher-order corrections that do not involve undetermined functions).

This, however, does not mean, that $\phi_{\rm R}$ is the same as in the leading order RGPE because the RG equation itself does change.
 
Indeed, the independence on renormalization scales can be ensured by demanding that 
\begin{align}
\partial_\mu A & = \varepsilon^2 A|A|^2 \partial_\mu \mathcal{S}_1+\varepsilon^4 A|A|^4 \Bigl(\partial_\mu \mathcal{S}_5 -\bigl(2\mathcal{S}_1 +\bar{\mathcal{S}}_1\bigr)\partial_\mu \mathcal{S}_1 \nonumber \\
& \phantom{=} - \partial_\mu\mathcal{Z}\Bigr)\,.
\end{align}

When we take the usual combination of differential consequences of the above RG equations, we obtain 
{\small \begin{gather}
2\I\partial_\theta A +\partial^2 A = \alpha \varepsilon^2 A |A|^2 
-\beta \varepsilon^4 A|A|^4-\varepsilon^4 A|A|^4 \mathcal{Y}\,, 
\end{gather}}
where
\begin{align}\label{eq:alphabeta1}
\alpha & \equiv \frac{10a_3^2}{3}+3a_4\,, \\ \label{eq:alphabeta2}
\beta & \equiv \frac{a_3^4}{18}+\frac{3a_4^2}{8}- 10a_6-\frac{56a_3a_5}{3}-9a_3^2a_4\,,
\end{align}
and where 
\begin{equation}
\mathcal{Y} = 2\I\partial_\theta \mathcal{Z} +\partial^2\mathcal{Z} - 2a_3 \bigl(\mathcal{S}_3+\mathcal{S}_2\bigr) + \frac{8}{3}a_3^2 \mathcal{S}_1+4a_3^2 \bar{\mathcal{S}}_1\,.
\end{equation}
As advertised, we now use the freedom in the definition of the dressed amplitude to remove all secular-term dependency in the above equation, namely, we choose such a $\mathcal{Z}$ so that $\mathcal{Y} = 0$.

This gives us a final RG equation in the form
\begin{equation}
2\I\partial_\theta A +\partial^2 A = \alpha \varepsilon^2 A |A|^2 
-\beta \varepsilon^4 A|A|^4\,.
\end{equation}
Let us now define a complex field
\begin{equation}
\Psi \equiv \varepsilon \sqrt{\frac{5 a_3^2}{6}+\frac{3a_4}{4}} A \e^{\I \theta}\,.
\end{equation}
The RG equation can be now reexpressed as
\begin{equation}
\bigl(\partial^2 +1\bigr) \Psi = 4 \Psi |\Psi|^2 - 16 \frac{\beta}{\alpha^2}\Psi |\Psi|^4\,.
\end{equation}
This equation is no-longer universal in the sense that it explicitly contain parameters of the model through definitions \refer{eq:alphabeta1}-\refer{eq:alphabeta2}.

A single $Q$-ball solution reads
\begin{equation}
\Psi^{\mbox{$Q$-ball}} = \frac{\lambda \e^{\I\omega t}}{\sqrt{\tilde \omega \cosh(2\lambda x)+1}}\,,
\end{equation}
where $\omega = \sqrt{1-\lambda^2}$ and where 
\begin{equation}
\tilde\omega = \sqrt{1-\frac{16\beta \lambda^2}{3\alpha^2}}\,.
\end{equation}

As an example, let us consider the sine-Gordon model, i.e., $a_3 = a_5 = 0$, $a_4 = 1/6$, and $a_6 = -1/120$. Plugging in the corresponding $Q$-ball solution
\begin{gather}
\Psi_{\rm sG}^{\mbox{$Q$-ball}} = \frac{\lambda \e^{\I\omega t}}{\sqrt{1+\sqrt{1-2\lambda^2}}\cosh(2\lambda x)}
\end{gather}
 into the renormalised solution truncated to the $\varepsilon^3$ terms outputs
 \begin{gather}
 \phi_{\rm R}^{\mbox{sG}} = \frac{\lambda \cos(\omega t)}{24 \bigl(1+\sqrt{1-2\lambda^2}\cosh(2x \lambda)\bigr)^{3/2}}\Bigl(48+\lambda^2\nonumber \\
 -2\lambda^2 \cos(2\omega t) +48 \sqrt{1-2\lambda^2}\cosh(2x\lambda)\Bigr)\,.
 \end{gather}
 Notice that this solution is only defined up to $\lambda \leq 1/\sqrt{2} \approx 0.71$.
 
 If we now make an expansion in powers of $\lambda$ that are not inside $\cos(\omega t)$ and $\cosh(\lambda x)$, we get 
\begin{equation}
\phi_{\rm R}^{\mbox{sG}} \approx \bigl(4\lambda+2\lambda^3\bigr)\frac{ \cos(\omega t)}{\cosh(\lambda x)}-\frac{4\lambda^3}{3}\frac{\cos(\omega t)^3}{\cosh(\lambda x)^3}\,.
\end{equation}
This precisely match the similar expansion of the breather solution 
\begin{equation}
\phi_{\rm breather} = 4\arctan\Bigl(\frac{\lambda \cos(\omega t)}{\sqrt{1-\lambda^2}\cosh(\lambda x)}\Bigr)\,,
\end{equation}
to $\lambda^3$.

As another example, let us consider a cubic model, i.e., $a_3 =1$ and $a_{4,5,6}= 0$. The corresponding $Q$-ball solution reads
\begin{equation}
\Psi_{\rm cubic}^{\mbox{$Q$-ball}}  = \frac{3\sqrt{2}\lambda \e^{\I\omega t}}{\sqrt{15+\sqrt{225-6\lambda^2}\cosh(2x \lambda)}}\,.
\end{equation}

If we supply this into the truncated renormalized solution and we perform similar expansion in $\lambda$'s as above, we obtain up to the third order
{\small \begin{gather}
\phi_{\rm R}^{\mbox{cubic}} \approx \sqrt{\frac{12}{5}}\lambda \frac{\cos(\omega t)}{\cosh(\lambda x)}+\frac{2\lambda^2}{5}\frac{3-\cos(2\omega t)}{\cosh^2(\lambda x)}\nonumber \\
+\sqrt{\frac{3}{5}}\frac{\lambda^3\cos(3\omega t)}{\cosh^3(\lambda x)}+\frac{\lambda^3\cos(\omega t)}{50\sqrt{15}\cosh(\lambda x)}\Bigl(2-\frac{1}{\cosh^2(\lambda x)}\Bigr)\,.
\end{gather} }
We remark that this is not identical to FFHL expansion, but there are similar terms with slightly different coefficients. 

In both of these examples, the higher order corrections are in practice invisible if plotted for sufficiently small $\lambda$, for which the renormalized solution matches well with the numerical one. Hence, they do not constitute a significant improvement in accuracy. Rather, they illustrate that the RGPE scheme is consistent at the higher orders.

\section{Summary}

In the present work, we have resolved the long-standing, important problems related to oscillons.\footnote{This work is also presented in a shortened, Letter version in Ref.~\cite{Letter}.}

First of all, we have established a {\it nontrivial}, RG-based, {\it relationship between oscillons and $Q$-balls}. An intimate connection between these objects, especially in the approach based on the so-called $I$-balls, has been conjectured for a long time. We have shown that this is indeed the case: oscillons and $Q$-balls are related, however, in a much more nontrivial way. The underlying, hidden $Q$-balls, are solutions of the RG equation for the complex amplitude and are seeds for the oscillons. They enter into a dressing formula that relates the $Q$-balls with the oscillons. 

Unexpectedly, the $Q$-ball equation enjoys a sort of {\it universality property}. This means that various scalar field theories in (1+1) dimensions, at least at the leading order, possess exactly the same RG-based $Q$-ball equation. This indicates that oscillons in theories with the same RG equations, as e.g., $\phi^3$, and various versions of $\phi^4$, should have similar properties and belong to the same universality class. This also explains why oscillons in the exotic $\phi^6$ are quite different  and why they do not reveal any amplitude modulations. Hence, our framework allows a natural way of {\it classification of oscillons}. 

In the generic universality class, defined by the complex $\phi^4$ equation, we have models with the cubic and/or quartic self-interaction terms in the potential. Higher terms may also exist. Oscillons of the $\phi^3$, the reverse $\phi^4$, and the double well $\phi^4$ and many other \cite{Pujo-2} belong to this class. Here the same single $Q$-ball solution generates an approximation which very well reproduces dynamics of the simplest, unexcited, i.e., unmodulated oscillons. This RG-based approximation is similar, but not identical to the standard FFHL expansion \cite{Fodor:2008es}. 

Importantly, in the sine-Gordon case, if higher-order terms in the RG expansion are taken into account we see a convergence of the renormalized solution to the breather. 

The novelty and strength of the RG-based approach are clearly visible in the context of the modulated oscillons. Due to the similarity of this RG equation to the complex sine-Gordon ($\mathbb{C}$sG) theory, one can use the multi $Q$-ball solutions of the $\mathbb{C}$sG model. This is a natural way of introducing a higher number of degrees of freedom which now are associated with a higher number of $Q$-balls. This is a great improvement in comparison with \cite{Fodor:2008es}, allowing for the approximated analytical description of the much larger class of oscillons, which includes modulated oscillons as well. 

As a consequence, we have clarified the old question of {\it the origin of the amplitude modulation}, which is the most characteristic feature of oscillons. An excited, i.e., modulated, oscillon is a bound state of two unmodulated oscillons originating in a two $Q$-ball solution. The modulation is an effect of a nonlinear superposition of these two oscillons each with its own DoF, i.e., the fundamental frequency. This confirms the recent findings presented in \cite{Blaschke:2024uec}. 

We underline that the analyzed models, which belong to the universality class emerging from the complex $|\Psi|^4$ model of Eq.~\refer{eq:lag1} are qualitatively very different. They possess sphalerons with ($\phi^3$) or without (the reverse $\phi^4$) shape mode. They may have (the double well $\phi^4$) or they may not have ($\phi^3$ and the reverse $\phi^4$) topological soliton. Nonetheless, the oscillons are captured by the same $Q$-balls. 

Other universality classes, e.g., the class of the potentials without the cube and the quartic terms, have oscillons with the very distinct properties. Namely, they do not possess any modulated structure. This is associated with the lack of the similarity between $Q$-ball equation and the integrable $\mathbb{C}$sG theory. Hence, there are no two $Q$-ball solutions. In any case, properties of (multi) $Q$-balls in the RG-derived $Q$-ball equations seem to play a significant role for the properties of oscillons. Undoubtedly, this should be further studied. 

\vspace*{0.2cm}

There are more directions in which the present work can be continued. 

First of all, the established relation between oscillons and (particular) $Q$-balls suggests that various $Q$-ball phenomena (like e..g, charge swapping \cite{PS1}, superradiance \cite{PS2} or negative radiation pressure \cite{Ciurla:2024ksm}) may possess oscillon counterparts. This also indicates that various $ Q$-ball-tailored methods may be easily applied to oscillons.  For example, a collective coordinate model capturing the modulated structure of oscillons may be done along the construction presented in \cite{Bowcock:2008dn}. This should significantly improve the previous approaches based on the sphaleron \cite{Manton:2023mdr} or the Derrick modes  \cite{JQ} along the Relativistic Moduli Space approach \cite{AMORW}.

Next, connection to the integrable $\mathbb{C}$sG should be further investigated as it may provide a new insight into the problem of the unexpected stability of oscillons. Moreover, it opens a path to the analysis of quantum oscillons, which contrary to common belief \cite{H} may be long living objects \cite{J}. 

One should also better understand oscillons in the non-generic equivalency classes. As we have shown they differ significantly from the generic oscillons. However, apart from that, there is very little known about them. 

Finally, although our work focused on (1+1) dimensional oscillons, there might be no fundamental obstacles in applying this framework to higher dimensions. 

\vspace*{0.2cm}

We believe that the identified intimate relation between oscillons and $Q$-balls, which in a natural way introduces notion of the universality and integrability, opens a new avenue for understanding oscillons. To some extent, it also unifies these objects. 

\vspace*{0.4cm}

\acknowledgments
F.B. acknowledges the institutional support of the Research Centre for Theoretical Physics and Astrophysics, Institute of Physics, Silesian University in Opava and the support of Institute of Experimental and Applied Physics in Czech Technical University in Prague.
This work has been supported by the grant no. SGS/24/2024 Astrophysical processes in strong gravitational and electromagnetic fields of compact object. KS acknowledges financial support from the Polish National Science Centre 
(Grant NCN 2021/43/D/ST2/01122).  AW acknowledges support from the Spanish Ministerio de Ciencia e Innovacion (MCIN) 
with funding from European Union NextGenerationEU (Grant No. PRTRC17.I1) and Consejeria de Educacion from JCyL through the QCAYLE project, as well as MCIN Project 1114 No. PID2020-113406GB-I0 and the grant PID2023-148409NB-I00 MTM. 


\end{document}